\documentclass[11pt, a4paper]{article}
\usepackage[tableposition=top]{caption}
\usepackage[onehalfspacing]{setspace}
\usepackage[top=70pt,bottom=50pt,left=100pt,right=100pt]{geometry}

\usepackage{fancyhdr}
\usepackage{color}
\usepackage{amssymb,amsmath,amsthm,bbm,mathrsfs,ulem}
\usepackage{latexsym} 
\usepackage{graphicx} 
\usepackage{verbatim}
\usepackage{nicefrac} 
\usepackage{hyperref}
\usepackage[latin1]{inputenc} 
\usepackage{doc}
\usepackage[dvipsnames, table]{xcolor}

\usepackage{natbib}
\bibpunct{(}{)}{;}{a}{,}{,} 

\graphicspath{{./}}

\newtheorem{theo}{Theorem}[section]
\newtheorem{coro}[theo]{Corollary}
\newtheorem{prop}[theo]{Proposition}
\newtheorem{defi}[theo]{Definition}
\newtheorem{lemm}[theo]{Lemma}

\newtheorem{algo}[theo]{Algorithm}

\newcommand{\ew}{\mathbb{E}}     
\newcommand{\va}{\mathbb{V}\!ar} 
\newcommand{\co}{\mathbb{C}ov} 
\newcommand{\pr}{\mathbb{P}}     
\newcommand{\rn}{\mathbb{R}}     
\newcommand{\nn}{\mathbb{N}}		

\title{Bivariate change point detection:\\
	joint detection of changes in expectation and variance}
\author{
	{MICHAEL MESSER}\\[1ex]
	{Institute of Statistics and Mathematical Methods in Economics}\\
	{Vienna University of Technology}\\
    }
\date{}

\begin{document}
	\maketitle

\begin{abstract}
	A method for change point detection is proposed. We consider a univariate sequence of independent random variables with piecewise constant expectation and variance, apart from which the distribution may vary periodically. We aim to detect change points in both expectation and variance.
	For that, we propose a statistical test for the null hypothesis of no change points and an algorithm for change point detection. Both are based on a bivariate moving sum approach that jointly evaluates the mean and the empirical variance. The joint consideration helps improve inference as compared to separate univariate approaches. We infer on the strength and the type of changes with confidence. Nonparametric methodology supports the analysis of diverse data. Additionally, a multi-scale approach addresses complex patterns in change points and effects. We demonstrate the performance through theoretical results and simulation studies. A companion \texttt{R}-package \texttt{jcp} (available on CRAN) is discussed. \\
	
\end{abstract}

\section{Introduction}

%

The paper contributes to the field of change point detection which provides methods for the detection of structural breaks \--- change points \--- in stochastic sequences. Change point detection finds application in many areas of research,   e.g., oceanographic sciences \citep{Killick2010}, neuroimaging \citep{Aston2012}, telecommunication \citep{Zhang2009}, DNA sequencing \citep{Braun2000}, econometrics \citep{Zeileis2010}, to name a few. Change point problems are studied extensively, thus, many aspects of statistical methodology are addressed, e.g., hypothesis testing, change point estimation, model complexity, computational feasibility, data structures, practical performance, etc.,~see for example  the textbooks \citet{basseville,Csorgo1997,brodsky,Chen2000,Brodsky2017} or the review articles \citet{aue2013, Jandhyala2013}.

We study independent univariate random variables (RVs) that have piecewise constant expectation ($\mu$) and variance ($\sigma^2$). The goal is to detect changes in both $\mu$ and $\sigma^2$.
Apart from constant first and second moments within each section, the distribution is allowed to vary periodically, allowing for variability in higher moments. This periodicity can improve the mimicking of real data as compared to i.i.d.~RVs, the latter of which are contained as a special case. We provide a nonparametric method for the detection of change points which may occur on multiple time scales. 
There is vast literature on all aspects mentioned. Regarding independence, we mention \citet{Csorgo1988,Gombay2002,Horvath2005} ($U$-statistics), \citet{Horvath2007,Holmes2013} (empirical processes), and \citet{Siegmund1988,Frick2014,Fang2020} (likelihood ratios), the last two tackling multiple change points. 
For the detection of changes in $\mu$ we mention the nonparametric methods of \citet{Wolfe1984, Horvath2008, Jaruskova2010}; or \citet{Dehling2013}, and in particular regarding change points on multiple time scales we refer to \citet{Spokoiny2009,fryz2014,matteson2014}. For the detection of changes in $\sigma^2$ we mention the articles of \citet{hsu1977, inclan1994,chen1997, whitcher2000,Killick2013,Korkas2017} and a paper for changes in scale by \citet{Gerstenberger2019}.
 Combining both, changes in $\mu$ and $\sigma^2$, we refer to \citet{Pein2016} who aim at detecting changes in $\mu$ by allowing $\sigma^2$ to change simultaneously. Vice versa, \citet{Gao2018} or \citet{Dette2015} study changes in $\sigma^2$ allowing $\mu$ to vary smoothly. 
Further, we note that \citet{Gorecki2018} and \citet{Messer2014} aim at detecting changes in $\mu$ (first moment) while a certain degree of heteroscedasticity, i.e., variability in the second moment, is allowed. In this work, the latter is extended in the sense that we detect changes in both $\mu$ and $\sigma^2$, while we also allow for variability in higher moments.

We propose a bivariate method to jointly quantify change in $\mu$ and $\sigma^2$. For that we consider two moving sum processes (MOSUM): data are pointwise restricted to adjacent windows from which first the means and second the empirical variances are compared. The first statistic is sensitive to changes in $\mu$ and the second to changes in $\sigma^2$. See Figure \ref{fig1} for an example of the two univariate processes (B,C) and their joint consideration (D), details are explained later. Regarding univariate MOSUM we mention  \citet{Steinebach1995,Antoch1999,Huskova2001} and \citet{Eichinger2018}.
Importantly, the bivariate approach presented here helps overcome flawed inference as compared to separate univariate approaches. Also, it enables a straightforward interpretation of the types of changes ($\mu$, $\sigma^2$ or both), as well as their strengths (effect sizes), while also controlling the error (confidence) of such statements. Methodologically, we construct first a test for the null hypothesis of an absence of change points. Second, we propose an algorithm for change point detection which is run if the null hypothesis is rejected. 

We mention three benefits: first, the proposed theory is highly nonparametric which enables wide-ranging applicability. Strong performance is shown under different distributional assumptions, including normal-, exponential- or gamma-distributed data. Second, multi-scale aspects are captured, meaning that the occurrence of fast as well as slow change points with different strengths of effects are tackled.
Third, the method is ready to use. Theory is proven with all unknown quantities replaced by appropriate estimators. A companion $\texttt{R}$-package $\texttt{jcp}$ (\textit{joint change point} detection, \citet{Messer2019}) is  provided whose graphical output facilitates interpretation. 

The paper is organized as follows: in Section \ref{sect:idea} we present basic concepts. In Section \ref{sect:model} we introduce the model, and in Section  \ref{sect:evj} we define the MOSUM statistics. We construct the test in Section \ref{sect:test} and discuss change point detection in Section \ref{sect:cpd}. In Section \ref{sect:jht_c} we give additional theory. In Section \ref{sect:practical} practical aspects are considered: methodology is extended, the \texttt{R}-package \texttt{jcp} is discussed, simulation studies are performed, and a real data example is shown. Proofs and auxiliary results are given in the Appendix. 


\section{The idea of testing and change point detection}\label{sect:idea}
\begin{figure}[h!]
	\hrule\vspace{0.5em}
	\includegraphics[width=0.5\textwidth,angle=0]{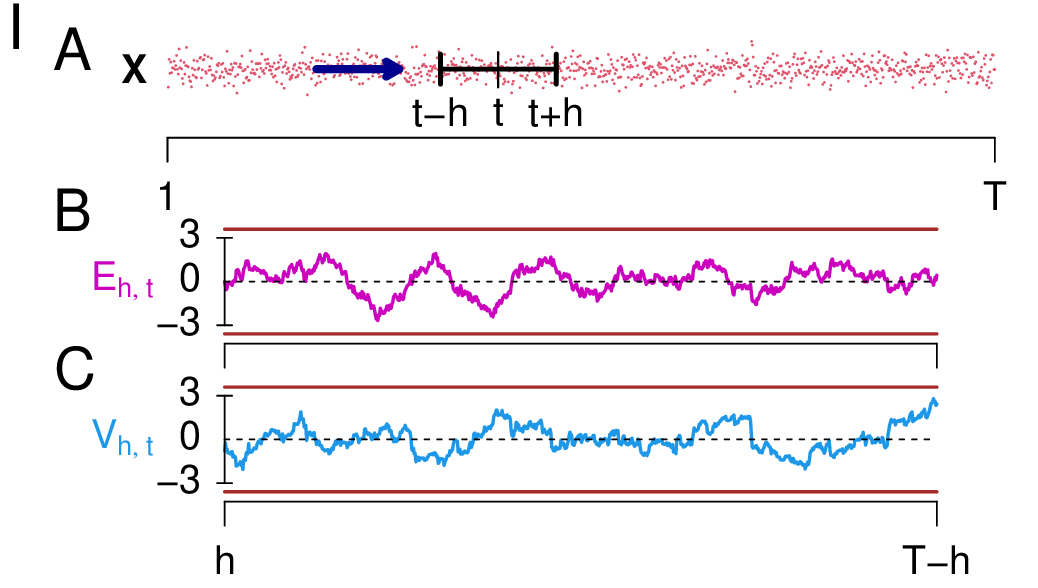}
	\includegraphics[width=0.5\textwidth,angle=0]{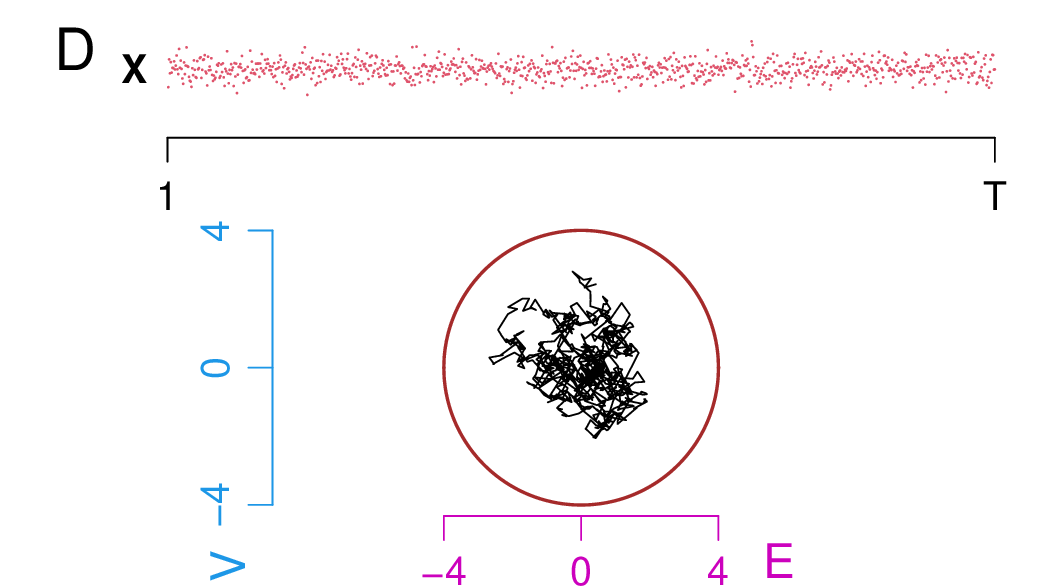}\\
	\vspace{0.5em}\hrule\vspace{0.5em}
	\includegraphics[width=0.5\textwidth,angle=0]{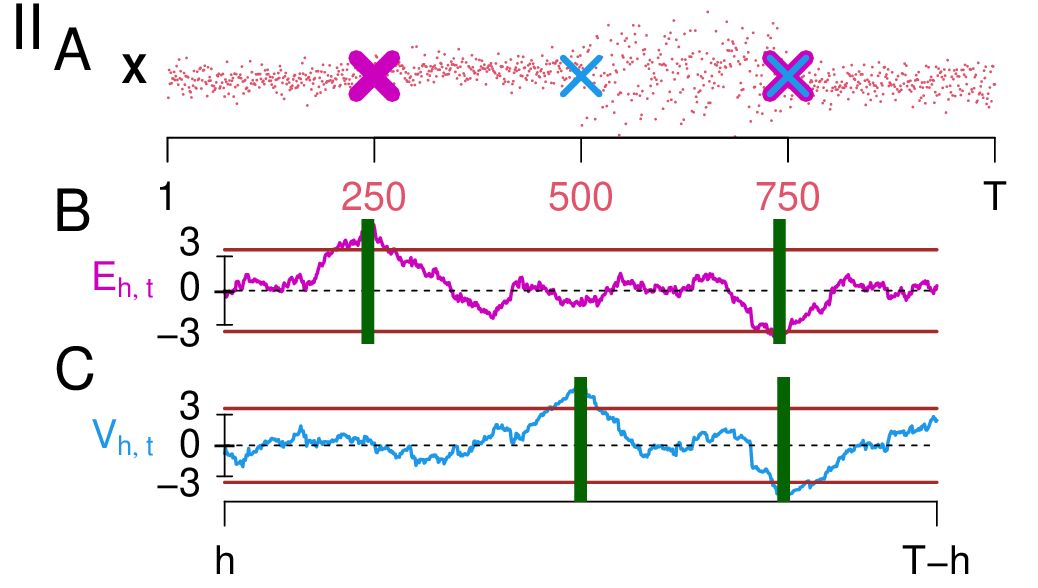}
	\includegraphics[width=0.5\textwidth,angle=0]{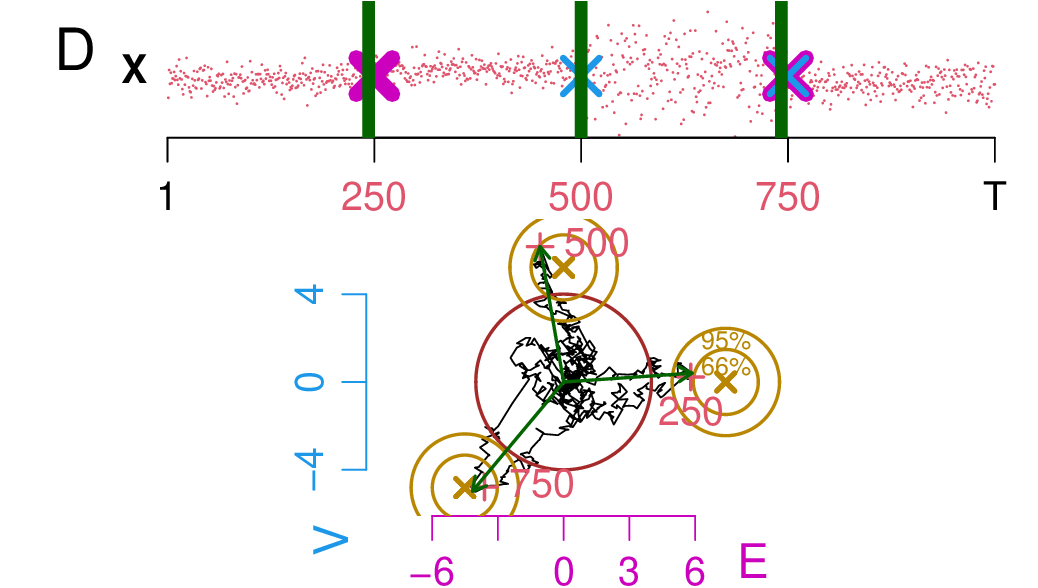}\\
	\vspace{0.5em}\hrule\vspace{0.5em}
	\includegraphics[width=0.5\textwidth,angle=0]{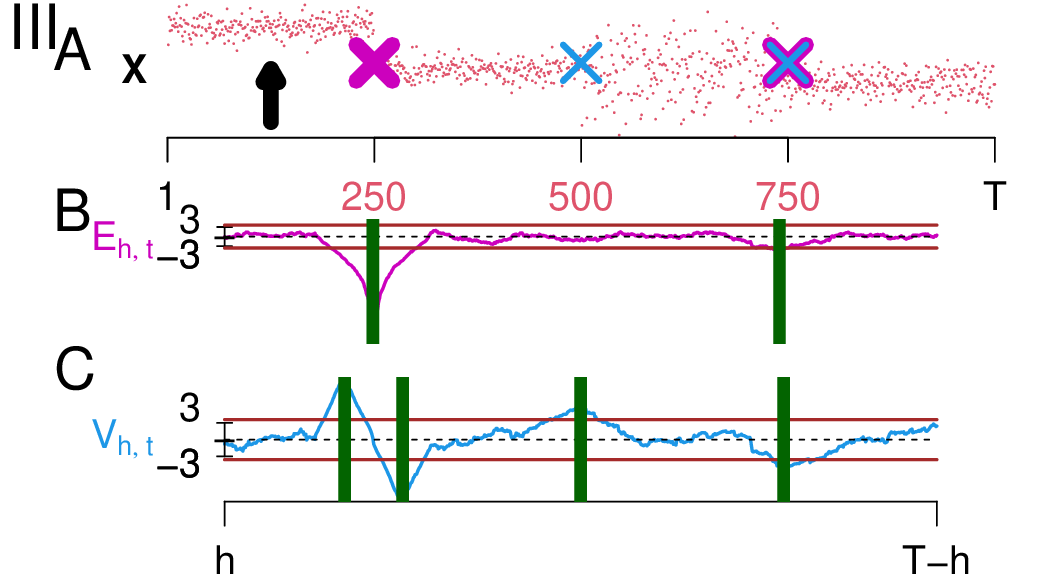}
	\includegraphics[width=0.5\textwidth,angle=0]{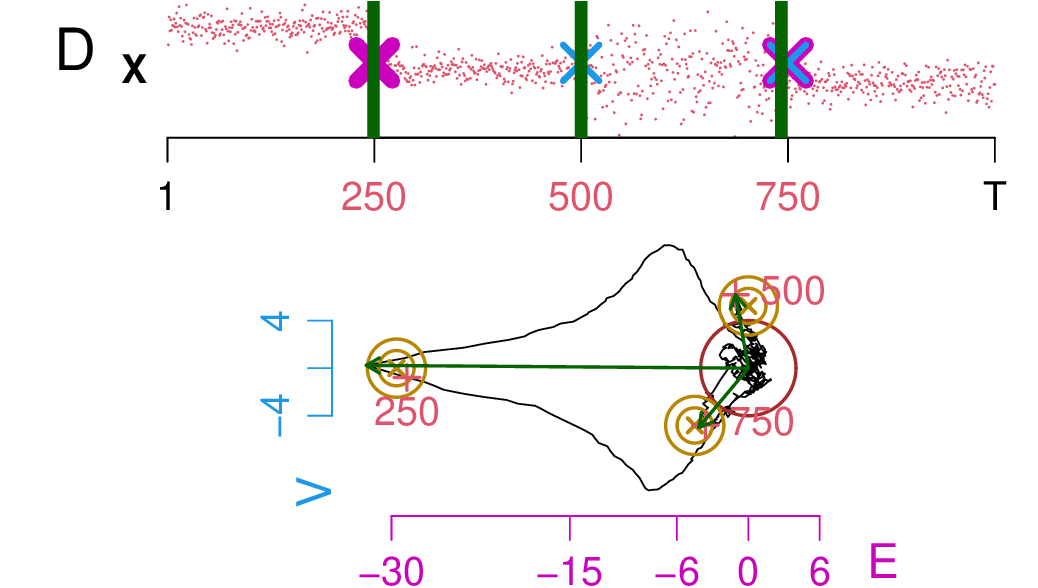}
	\caption{Univariate (B,C) and bivariate (D) procedures. $T=1000$, $\mathbf X$ piecewise i.i.d. $N(\mu,\sigma^2)$ distributed (A), I: $C=\emptyset$. II, III: $C=\{250,500,750\}$ (crosses), green bars mark $\hat C$. I: $\mu=10$, $\sigma=4$. II: $\mu =5,10,10,3$, $\sigma=4,4,12,5$, and $\hat C=\{243,740\}$ (B), $\hat C=\{500,745\}$ (C), $\hat C=\{243,500,742\}$ (D). III: $\mu=30,10,10,3$, $\sigma=4,4,12,5$, and $\hat C=\{249,740\}$ (B), $\hat C=\{215,285,500,745\}$ (C), $\hat C=\{249,500,742\}$ (D).}
	\label{fig1}
\end{figure}

We show how MOSUM processes are used for testing and change point detection. Particularly, we motivate the bivariate aspect. For that, consider the three processes in Figure \ref{fig1}, differentiated in panels I, II, and III.


We consider $T=1000$ independent RVs that are piecewise $N(\mu,\sigma^2)$ distributed, see red points in segments A and top D (coinciding). In panel I, there are no change points, $C=\emptyset$. In panels II and III there are three change points $C=\{250,500,750\}$. In both II and III the types of changes coincide: a change in $\mu$ at $250$ (purple cross), an increase in $\sigma^2$ at $500$ (blue cross), and a decrease in both $\mu$ and $\sigma^2$ at $750$ (purple and blue cross). II and III only differ at $250$: there is a small increase in $\mu$ in II and a prominent decrease in $\mu$ in III.

MOSUM processes are shown in segments B, C, and D. Two adjacent windows of size $h=70$ (see panel IA) are shifted through time and statistics are evaluated locally from the RVs in the windows. The first statistics $E_{h,t}$ (B, magenta) is Student's unpooled $t$-statistic, i.e., Welch's statistic, which compares the empirical means, see (\ref{EandV}). It is sensitive to changes in $\mu$. The second statistic $V_{h,t}$ (C, blue) similarly compares the emprical variances and is thus sensitive to changes in $\sigma^2$. In segment D we see a bivariate process (black) given by the joint statistic $J_{h,t}:=(E_{h,t},V_{h,t})^\mathsf{T}$, where $\mathsf{T}$ denotes transposing. If the windows do not overlap a change point, then the two estimates from the left and right window typically resemble each other, resulting in a statistic close to zero. A strong deviation from zero indicates a change.
We see brown rejection boundaries (B, C horizontal lines at $\approx \pm 3.59$, D circle with radius $\approx 4$. The boundaries coincide between the panels but appear different due to the scaling of axes).  The idea is that the processes entirely lie within the boundaries with a predefined probability of $1-\alpha$, here $95\%$, in case null hypothesis $\mathcal H_0:C=\emptyset$ holds true, see panel I and Section \ref{sect:test} for details. 
If the boundaries are crossed at some point (II and III in B,C and D), the null hypothesis is rejected. After rejection, the set $C$ is estimated (green bars) via successive $\arg\!\max$ estimation: find the largest deviation form zero, take the argument as a change point estimate, delete the process in the $h$-neighborhood of the estimate and repeat until the remaining process lies within the boundaries (also see Figure \ref{fig3}).  In Figure \ref{fig1} panel II inference is reasonable as in B the two changes in $\mu$, in segment C the two changes in $\sigma^2$, and in D all three change points were detected. In III expectation change detection succeeds in B, but unfortunately variance inference in segment C fails as two changes in $\sigma^2$ are falsely estimated. This is caused by the prominent change in $\mu$. For an intuition, recall that the empirical variance is the mean squared deviation from the mean. Thus, when $\mu$ changes, this impacts the mean and thus the variance. This problem also appears in panel II, but it is practically negligible as the change in $\mu$ is small. Vice versa, changes in $\sigma^2$ practically do not impact expectation change detection, intuitively, because the mean is the first moment and not affected if only the second moment changes. Statements about robustness are subject of Section \ref{sect:jht_c}. 
One way to tackle false $\sigma^2$-inference is to incorporate information about $\mu$: if the empirical variances are centered correctly then $\sigma^2$-estimation will not systematically react falsely to changes in $\mu$. In the context of stochastic point processes this was studied in \citet{Albert2017}. A second way of treating this problem is presented in this paper: the joint observation of both processes. In panel IIID the bivariate MOSUM overcomes the problem and successfully detected three change points. 

The golden dartboards in segment D show the asymptotic distribution of $J_{h,c}$ at a true change point $c$. This distribution is bivariate normal, the golden cross marks the expectation and the circles the $66\%$- and $95\%$-contour lines, see Proposition \ref{prop:jhc}.
When the sliding windows run into $c$, then $(J_{h,t})_t$ starts an excursion approaching the center of the dartboard. The red crosses ($+$) mark $J_{h,c}$ i.e., they can be considered a realization from a dartboard. When the sliding windows trespass $c$, the process returns to zero fluctuation.  Glimpse at Figure \ref{figshark} for the systematics of the excursions. 
The type and the strength of the change affect the systematic aspect of the excursion: if there is a change only in $\mu$, the process tends to leave the circle in horizontal direction, to the right if there is an increase in $\mu$ and to the left in case of a decrease. Indeed, at time $250$ the dartboard is shifted only along the abscissa. In panel II it is found at the right side as there is an increase in $\mu$, and in III it is shifted left due to the decrease in $\mu$. Further, in panel III it lies far out as the change in $\mu$ is strong. If there is a change only in $\sigma^2$, the process systematically moves along the ordinate, upwards in case of an increase and downwards in case of decrease. At $500$ the dartboard is shifted only along the ordinate. In case of a change in both $\mu$ and $\sigma^2$, the process leaves the circle in both horizontal and vertical direction. Indeed, at $750$ the dartboard lies south-west due to the decline in both $\mu$ and $\sigma^2$. Vice versa, the location of a dartboard facilitates change point interpretation, see Section \ref{sect:cpd}.

%

\section{The Model}\label{sect:model}
\begin{figure}[h!]
	\centering  
	\includegraphics[width=0.6\textwidth,angle=0]{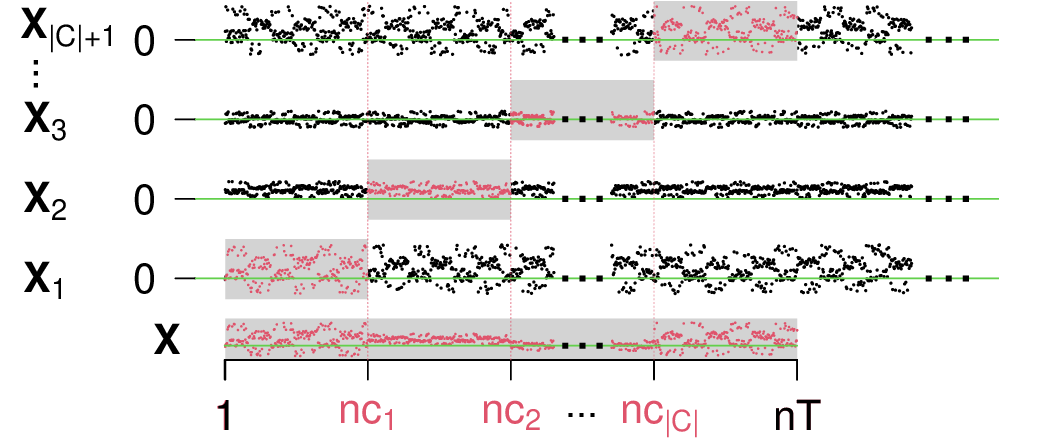}
	\caption{Construction of $\mathbf{X}$. All
		$\textbf{X}_1,\textbf{X}_2,\textbf{X}_3$ and $\textbf{X}_{|C|+1}$ with $p=80$ derive from mixtures of uniform RVs (see previous paragraph) with $\mu=10,10,0,10$ and $\sigma=12,4,4,12$.
	}
	\label{fig7}
\end{figure}

\paragraph{Auxiliary processes}
Consider a single probability space $(\Omega,\mathcal A,\pr)$ throughout. 
\begin{defi}\label{defi:aux}
\normalfont
A sequence $\mathbf X=(X_i)_{i=1,2,\ldots}$ of independent RVs of $\mathscr L^4(\Omega,\mathcal A,\pr)$ is called an \textit{auxiliary process}, if there exists a $p\in\{1,2,\ldots\}$ such that for all $m\in \{1,2,\ldots, p\}$ it holds that
$X_{ip+m} \stackrel{d}{=}X_{m}$ for all $i=1,2\ldots$, and also that
\begin{align}\label{mo12}
\ew[X_m] = \ew[X_1] =:\mu   && \textrm{and} && \va(X_m) = \va(X_1) =:\sigma^2>0. 
\end{align}
\end{defi}
The distribution varies with period $p$ while the first two moments are constant. 
This includes i.i.d.~sequences. Higher moments may vary periodically. Thus, for $k=1,\ldots,4$ we abbreviate the $k$-th moment and centered moment via
$\mu(m)^{\langle k \rangle}  := \ew[X_m^k]$ and $\mu(m)^{\{k\}} := \ew[(X_m-\mu)^k]$. We set averaged moments within a period 
\begin{align}\label{theo_moments}
\mu^{\langle k \rangle} := \frac{1}{p} \sum\nolimits_{m=1}^p \mu(m)^{\langle k \rangle}
&& \textrm{and} &&
\mu^{\{k\}} := \frac{1}{p} \sum\nolimits_{m=1}^p \mu(m)^{\{k\}}.
\end{align}
For $k=1,2$, (\ref{mo12}) implies constants $\mu(m)^{\langle 1 \rangle}=\mu$, $\mu(m)^{\langle 2 \rangle} = \mu^{\langle 2 \rangle}$, $\mu(m)^{\{1\}}=0$ and $\mu(m)^{\{2\}}=\sigma^2$. Further set $\nu(m)^2:=\va((X_m-\mu)^2)=\mu(m)^{\{4\}}-\sigma^4$ and the average
\begin{align}\label{nu}
\nu^2:=\frac{1}{p}\sum\nolimits_{m=1}^p \nu(m)^2 = \mu^{\{4\}}-\sigma^4,
 && \textrm{and also set} && \rho := \frac{\mu^{\{3\}}}{\sigma\cdot \nu}.
\end{align}
Note that $\sigma^2>0$ implies $\nu(m)^2>0$ and thus $\nu^2>0$. Further note  $\mu(m)^{\{3\}}=\co(X_m,(X_m-\mu)^2)$. It is $|\rho|<1$ as $(X_m-\mu)^2$ is not linear in $X_m$. 
Call the averages $\mu$, $\sigma^2$, $\nu^2$, $\rho$, $\mu^{\{k\}}$ and $\mu^{\langle k \rangle}$ in (\ref{mo12}), (\ref{theo_moments}) and (\ref{nu}) the population parameters. 
Condition (\ref{mo12}) is sufficient for moment estimators to appropriately estimate all population parameters \--- the periodicity is averaged out over large samples.\\ 
For $d=1,2,\dots$ let $N_d(w,\Sigma)$ denote the $d$-variate normal distribution with expectation $w\in\rn^d$ and $d\times d$ covariance matrix $\Sigma$.  
For the mean $\hat\mu :=(1/n)\sum_{i=1}^n X_i$ and the empirical variance $\hat \sigma^2:=(1/n)\sum_{i=1}^n(X_i-\hat \mu)^2$ we jointly obtain as $n\to\infty$
\begin{align}\label{asymp_mean_var}
\sqrt{n}
\left[
\begin{pmatrix}
\hat{\mu}\\
\hat\sigma^2
\end{pmatrix}
-
\begin{pmatrix}
\mu\\
\sigma^2
\end{pmatrix}
\right]
\stackrel{d}{\longrightarrow}
N_2(0,\Sigma)
\qquad 
\textrm{with}
\qquad
\Sigma :=
\begin{pmatrix}
\sigma^2 & \rho\sigma\nu\\
\rho\sigma\nu & \nu^2
\end{pmatrix},
\end{align}
while $\stackrel{d}{\longrightarrow}$ denotes convergence in distribution, see Proposition \ref{prop:jhc}. $\Sigma$ is regular as $det(\Sigma)=(1-\rho)\sigma^2\nu^2>0$.\\ 
We give an example: let $Y$ be a mix of uniform RVs with density $f_Y(y)=(3/(4d))\mathbbm{1}_{[\mu-d,\mu]}(y)+(1/(4d))\mathbbm{1}_{[\mu+d,\mu+2d]}(y)$ for $\mu\in \rn$ and $d>0$. Then $\ew[Y]=\mu$ and $\va(Y)=5d^2/6=:\sigma^2$, and $f_Y$ has positive skewness as $3>1$. Further, let $Z$ have density $f_Z(z)=f_Y(2\mu-z)$, i.e., $f_Z$ results from $f_Y$ by mirroring at $\mu$. Thus, also $\ew[Z]=\mu$ and $\va(Z)=\sigma^2$, but skewness switches the sign. Consider independent copies of $Y$ and $Z$. Set $p$ even. Repeatedly let $p/2$ copies of $Z$ follow $p/2$ of $Y$. This is an auxiliary process with period $p$, see Figure \ref{fig7} $\mathbf{X}_1,\ldots,\mathbf{X}_{|C|+1}$.

\paragraph{The Model $\mathcal M$}
A piecewise combination of auxiliary processes constitutes a process with change points: we fix $T\in\nn\backslash\{0,1\}$ and consider a subset $C \subset \{1,\ldots,T-1\}$ of cardinality $|C|$ with ordered elements $c_1 < c_2< \ldots < c_{|C|}$. Call $c_u$ the $u$-th change point and $C$ the set of change points.
Given $C$, we consider $|C|+1$ independent auxiliary processes $\mathbf X_1,\ldots,\mathbf X_{|C|+1}$, with $\mathbf X_u=(X_{u,i})_{i=1,2,\ldots}$ and population parameters indexed by $u$, e.g., $\mu_u:=\ew[X_{u,1}]$ and $\sigma_{u}^2:=\va(X_{u,1})$, for $u=1,\ldots,|C|+1$. We assume $(\mu_{u},\sigma_{u}^2) \not=(\mu_{u+1},\sigma_{u+1}^2)$, for $u=1,\ldots,|C|$, i.e., at any $c_u$ either the expectation or the variance (or both) change, while other parameters may change too. 
For asymptotics we let $T$ and $c_1,\ldots,c_{|C|}$ depend on a factor $n$, i.e., we switch to $nT$ and $nc_1,\ldots,nc_{|C|}$. For $n=1,2,\ldots$ we define a compound process
\[
X_{1,1},\ldots,X_{1,nc_1},X_{2,nc_1+1},\ldots,X_{2,nc_2},\ldots, X_{|C|+1,nc_{|C|}+1},\ldots,X_{|C|+1,nT},
\]
i.e., after $nc_u$ we enter $\mathbf X_{u+1}$ with $(\mu_{u+1},\sigma_{u+1}^2)$. 
Given $T$, the set $\mathcal M$ of such processes constitutes the model. For asymptotics we let $n\to\infty$ and treat $C$, as well as all population parameters including the periods, as fixed, i.e., not depending on $n$, but unknown. By increasing $n$, the time we remain in each $\textbf{X}_{u}$ increases linearly, constituting a triangular setup. 
We call $n=1$ the real time scenario, see Figure \ref{fig7}.


For $\mathbf{X}\in\mathcal M$ we want to test $\mathcal H_0: C=\emptyset$. In case of rejection we aim at estimating  $C$. If $C=\emptyset$ we omit indices $\mu:=\mu_1$, $\sigma^2:=\sigma_1^2$, $\nu^2:=\nu_1^2$ etc.


\section{The moving sum processes}\label{sect:evj}
For $\textbf{X}\in\mathcal M$ we study moving sum processes in which we locally evaluate the RVs restricted to a \textit{left} window $(t-h,t]$ and an  adjacent \textit{right} window $(t,t+h]$. A comparison of means is sensitive to changes in the expectation and a comparison of empirical variances is sensitive to changes in the variance. 

\paragraph{Definition of $(E_{h,t}^{(n)})_t$ and $(V_{h,t}^{(n)})_t$ via local parameter estimators}
Let $h\in\{2,3,\ldots,\lfloor T/2 \rfloor\}$ be a window size independent of $n$, while $\lfloor \cdot \rfloor$ denotes the floor function, and let $t\in[h,T-h]$. For $n=1,2,\ldots$ set
\begin{align}\label{EandV}
	E_{h,t}^{(n)} := \frac{\hat{\mu}_r- \hat{\mu}_\ell}{[(\hat\sigma_r^2 + \hat\sigma_\ell^2)/(nh)]^{1/2}} 
	\qquad\textrm{and}\qquad
	V_{h,t}^{(n)} := \frac{\hat\sigma_r^2- \hat\sigma_\ell^2}{[(\hat\nu_r^2 + \hat\nu_\ell^2)/(nh)]^{1/2}}.
\end{align}
We define the estimators in (\ref{EandV}). The subscripts $\ell$ and $r$ indicate local evaluation, while $I_\ell:=\{\lfloor n(t-h)\rfloor+1,\ldots,\lfloor nt\rfloor \}$ and $I_r:=\{\lfloor nt\rfloor+1,\ldots,\lfloor n(t+h)\rfloor \}$ are the indices associated with the windows. We strengthen the dependence on $n$, $t$ and $h$, which is inherited to the estimators below, but omitted for simplicity. For $j\in\{\ell,r\}$ and $k\in\{1,\ldots,4\}$ we define estimators for the moments in (\ref{theo_moments}) and (\ref{nu}) via 
\begin{align}\label{moments}
	\hat\mu_j^{\langle k \rangle} := 
	\frac{1}{nh}
	\sum\nolimits_{i\in I_j} X_i^k
	\qquad\textrm{and}\qquad
	\hat\mu_j^{\{k\}} :=
	\frac{1}{nh}
	\sum\nolimits_{i\in I_j} (X_i-\hat\mu_j^{\langle 1\rangle})^k.
\end{align}
and further $\hat{\mu}_j := \hat{\mu}_j^{\langle 1\rangle}$, $\hat \sigma_j^2 := \hat\mu_j^{\{2\}}$ and $\hat\nu_j^2 := \hat\mu_j^{\{4\}}  - \hat\sigma_j^4$.

We study statistics in function space. For an interval $[a,b]\subset\rn_0$ let  $(\mathcal D_{\rn^d} [a,b],d_{SK})$ and $(\mathcal D_{\rn^d} [a,b],\|\cdot\|)$ denote the spaces of $\rn^d$-valued c\`adl\`ag-functions on $[a,b]$ equipped with Skorokhod topology $d_{SK}$ or the supremum norm $\|\cdot\|$. Convergence w.r.t.~$\|\cdot\|$ implies convergence regarding $d_{SK}$. Analogously, for $[0,\infty)$ consider $(\mathcal D_{\rn^d} [0,\infty),d_{SK})$. For $n=1,2,\ldots$ the estimators in (\ref{moments}) and thus $(E_{h,t}^{(n)})_t$ and $(V_{h,t}^{(n)})_t$ constitute processes in $(\mathcal D_\rn[h,T-h],d_{SK})$. 
Omitting the superscript $n$ abbreviates $n=1$, e.g., $E_{h,t}:=E_{h,t}^{(1)}$.
See Figure \ref{fig1} for $(E_{h,t})_t$ (B, magenta) and $(V_{h,t})_t$ (C, blue) evaluated from $\bold X$ (A, red). 
If $C=\emptyset$ then the estimators are functionally strongly consistent for their population parameters:
\begin{lemm}\label{conv:est}
	Let $\mathbf X\in\mathcal M$ with $C=\emptyset$. For $j\in\{\ell,r\}$ it holds in $(\mathcal D_{\rn}[h,T-h],d_{\|\cdot\|})$ as $n\to\infty$ almost surely that
	$(\hat\mu_j^{\langle k\rangle})_t\to(\mu^{\langle k\rangle})_t$ and 
	$(\hat\mu_j^{\{k\}})_t\to(\mu^{\{k\}})_t$, for $k\in\{1,\ldots,4\}$.
\end{lemm}	
\noindent
Consequently, $(\hat\sigma_j^2)_t\to (\sigma^2)_t$ and $(\hat\nu_j^2)_t\to (\nu^2)_t$ as $n\to\infty$ almost surely (a.s.).

\paragraph{The joint process $(J_{h,t}^{(n)})_t$}
We consider the joint process  $(J_{h,t}^{(n)})_t$ via $J_{h,t}^{(n)}:=(E_{h,t}^{(n)},V_{h,t}^{(n)})^\mathsf{T}$ in $(\mathcal D_{\rn^2}[h,T-h],d_{SK})$, see Figure \ref{fig1}D. Weak convergence to a bivariate Gaussian process is shown under $C=\emptyset$, see Proposition \ref{prop_main} below. From this the test is constructed in Section \ref{sect:test}. Further, convergences are extended to $C\not=\emptyset$ in Propositions  \ref{prop:jhc} and \ref{prop_main2}, supporting change point detection in Section \ref{sect:cpd}.

In the remainder of this section let $C=\emptyset$. In general, the components of $(J_{h,t}^{(n)})_t$ are correlated as both rely on $\textbf X$: in Figure \ref{fig2}A $(J_{h,t})_t$ primarily varies along the diagonal $x=y$. 
Symmetry of the RVs $\rho=\mu^{\{3\}}/(\sigma\nu)=0$ (see (\ref{nu})) is necessary and sufficient for the components to be asymptotically independent. Vice versa, skewness $\rho\not=0$ results in correlated components. We abusively speak about symmetry and skewness, meaning the average $\mu^{\{3\}}=(1/p)\sum_{m=1}^p \mu(m)^{\{3\}}$ to be $= 0$ or $\not=0$.

Skewness is captured in the correlation matrix 
\begin{small}
\begin{align}\label{gamma}
\Gamma :=
	\begin{pmatrix}
		1 & \rho\\
		\rho & 1
		\end{pmatrix}
=
 A D A^t :=
\begin{pmatrix}
1/\sqrt{2} & 1/\sqrt{2}\\
1/\sqrt{2} & -1/\sqrt{2}
\end{pmatrix}
\cdot
\begin{pmatrix}
1+\rho & 0\\
0 & 1-\rho
\end{pmatrix}
\cdot
\begin{pmatrix}
1/\sqrt{2} & 1/\sqrt{2}\\
1/\sqrt{2} & -1/\sqrt{2}
\end{pmatrix}.
\end{align}
\end{small}
$\Gamma$ is regular as $|\rho|<1$. The eigenvalue decomposition (\ref{gamma}) yields  $\Gamma^{1/2}$ and $\Gamma^{-1/2}$.
Unit diagonals of $\Gamma$ imply that the columns of $A$ span the diagonals $x=y$ and $x=-y$ in $\rn^2$. 
Symmetry  $\rho=0$ means $\Gamma = I = \Gamma^{1/2} = \Gamma^{-1/2}$, with identity $I$.
For an example of a skewed distribution consider $gamma(s,\lambda)$ with shape $s>0$ and rate $\lambda>0$, see also Figure \ref{fig2}A. We obtain $\sigma^2=s/\lambda^2$, $\nu^2 = 2[s^2+3s]/\lambda^4$ and $\mu^{\{3\}}=2s/\lambda^3$, and $\rho = (2/[s+3])^{1/2}>0$ depending only on $s$.
\begin{prop}\label{prop_main}
	Let $\mathbf X\in\mathcal M$ with $C=\emptyset$. In $(\mathcal D_{\rn^2}[h,T-h],d_{SK})$ it holds as $n\to\infty$ that 
		$(\Gamma^{-1/2} \cdot J_{h,t}^{(n)})_t
		\stackrel{d}{\longrightarrow}
		(\mathcal L_{h,t})_t$.
\end{prop}	

The bivariate limit process $(\mathcal L_{h,t})_{t}$ is given via
\begin{align}\label{limit}
\mathcal L_{h,t} :=
\left(
\frac{(W_{t+h} - W_t) - (W_t -W_{t-h})}{\sqrt{2h}}, 
\frac{(\mathcal W_{t+h} - \mathcal W_t) - (\mathcal W_t -\mathcal W_{t-h})}{\sqrt{2h}}
\right)^\mathsf{T},
\end{align}
while $(W,\mathcal W) :=(W_t,\mathcal W_t)_{t\ge 0}$ denotes a planar Brownian motion. The double windows are preserved. $(\mathcal L_{h,t})_{t}$ is a continuous $2h$-dependent bivariate Gaussian process that is isotropic. It is $\mathcal L_{h,t}\sim N_2(0,I)$ for all $t$.


\section{The statistical test}\label{sect:test}
\begin{figure}[h!]
	\centering  
	\includegraphics[width=0.5\textwidth,angle=0]{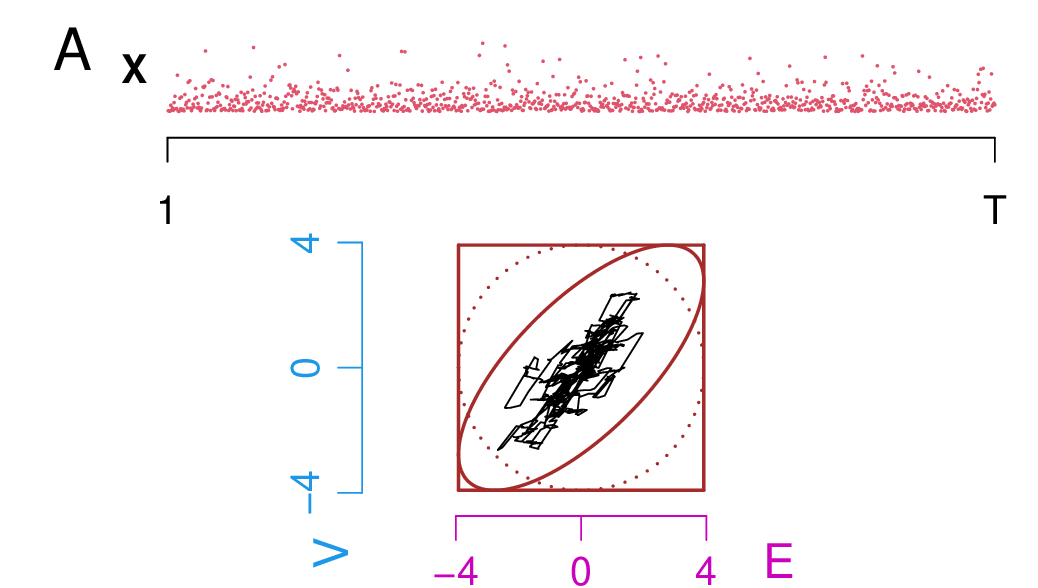}
	\includegraphics[width=0.5\textwidth,angle=0]{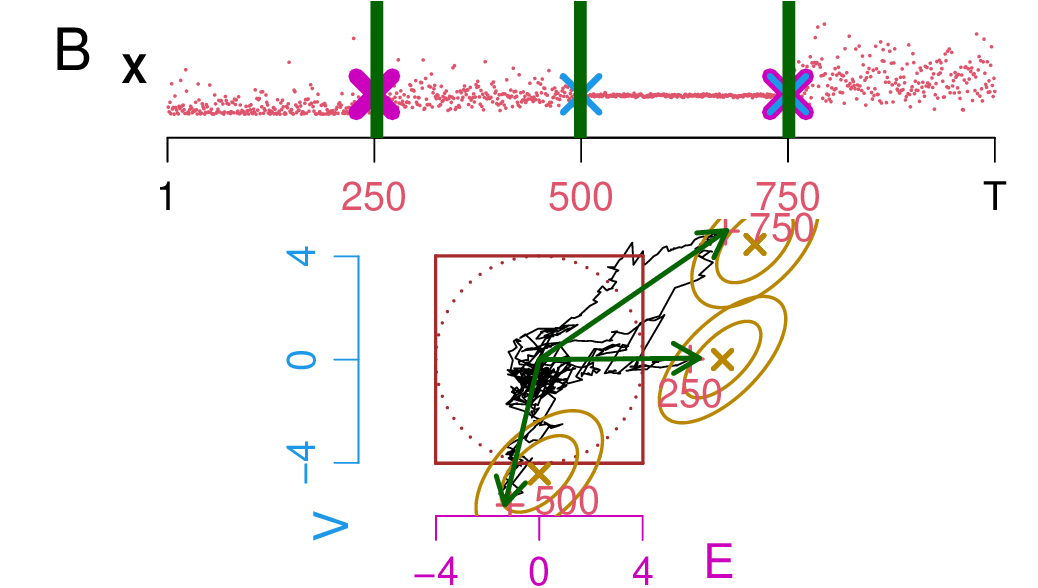}
	\caption{The bivariate procedure for skewed RVs. Top: $\mathbf X$ piecewise i.i.d.~$gamma(s,\lambda)$ distributed. Bottom: $(J_{h,t})_t$ (black), $\mathscr R$-boundaries (brown, $\mathscr C$ dotted, $\mathscr E,\mathscr S$ solid). A: $C=\emptyset$ and $gamma(1,1)=exp(1)$. $J_{h,t}\in\mathscr E (\subset\mathscr S$) for all $t$ and thus no rejection of $C=\emptyset$. B: $C=\{250,500,750\}$, $\mu=0.8,2,2,4$, $\sigma=1,1,0.1,2$. Rejection of $C=\emptyset$ as $(J_{h,t})_t$ leaves $\mathscr S$. $\hat C=\{253,499,751\}$ (green bars). '+' marks $J_{h,c}$ for $c\in C$ (red), asymptotic distribution of $J_{h,c}^{(n)}$ (golden ellipses, $66\%$- and $95\%$-contour lines).
	}
	\label{fig2}
\end{figure}

We test $\mathcal H_0: C=\emptyset$. A large deviation of $(J_{h,t}^{(n)})_t$ from zero speaks against $C=\emptyset$. We apply Proposition \ref{prop_main} to derive a rejection boundary from $(\mathcal L_{h,t})_t$ in simulations. Symmetry implies asymptotic isotropy of $(J_{h,t}^{(n)})_t$ and we set the boundary as a circle, see Figure \ref{fig1}. Generally, for possibly skewed RVs, $(J_{h,t}^{(n)})_t$ primarily varies along $x=y$ or $x=-y$, resulting in a rejection ellipse or square, see Figure \ref{fig2}.


\paragraph{Derivation of the test} 
Let $C=\emptyset$. For $(x,y)^\mathsf{T}\in\rn^2$ let $d_I((x,y)^\mathsf{T}) = (x^2 + y^2)^{1/2}$ denote the Euclidean and $d_\Gamma((x,y)^\mathsf{T})=[(x,y)\cdot \Gamma^{-1} \cdot(x,y)^\mathsf{T}]^{1/2}$ the Mahalanobis distance. Noting $d_I(\Gamma^{-1/2} \cdot J_{h,t}^{(n)}) = d_\Gamma( J_{h,t}^{(n)})$, Proposition \ref{prop_main} and continuous mapping imply convergence of the maximum, as $n\to\infty$ 
\begin{align}\label{Mh}
	M_h^{(n)}(\Gamma):=
	\max_{t\in[h,T-h]} d_\Gamma(J_{h,t}^{(n)})
	\stackrel{d}{\longrightarrow}
	\max_{t\in[h,T-h]} d_I(\mathcal L_{h,t}),
\end{align}
We use $M_h(\Gamma):=M_h^{(1)}(\Gamma)$ as a test statistic. We reject iff $M_h(\Gamma)$ exceeds the $(1-\alpha)$-quantile $Q$ of the limit distribution in (\ref{Mh}), given $\alpha\in (0,1)$. To the best of our knowledge there is no closed formula available. We choose $Q$ as a quantile of the approximated distribution derived in Monte Carlo simulations. 

In a similar setup \citet{Jaruskova2011} derived tail approximations for functionals of Brownian bridges to adjust $Q$. Here, $[d_I^2(\mathcal L_{h,t})]_t$ constitutes a $\chi^2$-process, which should allow to derive tail bounds as well, see \citet{Albin1990,Lindgren1980,Adler1990,Talagrand2014}. We mention high accuracy of simulations.

Equivalently, instead of comparing the maximum $M_h(\Gamma)$ to $Q\in\rn$, we can judge the entire process $(J_{h,t})_t$ w.r.t.~a rejection area $\mathscr R\subset\rn^2$. For that, define an ellipse 
$\mathscr E := \{(x,y)^\mathsf{T}\in\rn^2\; |\; d_\Gamma((x,y)^\mathsf{T}) \le Q \}$, see  Figure \ref{fig2}A.
$\mathcal H_0$ is rejected iff $(J_{h,t})_t$ enters $\mathscr R=\mathscr E^c$ at any time, while the superscript $c$ indicates the complement in $\rn^2$. In case of symmetry $\Gamma=I$, $\mathscr E$ equals a circle $\mathscr C$ with radius $Q$, i.e., 
$\mathscr C := \{(x,y)^\mathsf{T}\in\rn^2\; |\; d_I((x,y)^\mathsf{T}) \le Q \},$
thus $\mathscr R=\mathscr C^c$,
see Figure \ref{fig1}. For the skewed case, $\mathscr E$ derives from $\mathscr C$ by squeezing, $\mathscr E = \Gamma^{1/2}\cdot\mathscr C = \{\Gamma^{1/2} \cdot(x,y)^\mathsf{T} \; |\; (x,y)^\mathsf{T}\in \mathscr C \}$.

\paragraph{Treatment of the unknown correlation $\rho$}  $M_h^{(n)}(\Gamma)$ depends on unknown $\rho=\mu^{\{3\}}/(\sigma\nu)$. We propose three ways to cope with $\rho$ in practice. First, the symmetry assumption $\rho=0$ yields $M_h^{(n)}(I)$. Second, we consistently estimate $\rho$ locally by 
\begin{align}\label{hat_rho}
	\hat\rho := 
	\frac{\hat\mu_r^{\{ 3 \}} + \hat\mu_\ell^{\{3\}}}{(\hat\sigma_r^2 + \hat\sigma_\ell^2)^{1/2}\cdot(\hat\nu_r^2 + \hat\nu_\ell^2)^{1/2}},
\end{align}
recall (\ref{moments}). 
We define $\hat\Gamma$ by replacing $\rho$ with $\hat\rho$ in (\ref{gamma}), and consider  
the statistic
\begin{align}\label{M_gamma}
M_h^{(n)}(\hat\Gamma):=\max_t d_{\hat\Gamma}(J_{h,t}^{(n)}).
\end{align}
Convergence of $M_h^{(n)}(\hat\Gamma)$ as in (\ref{Mh}) holds true because in $(D_\rn[h,T-h],d_{SK})$ it is $(\hat \rho)_t\to (\rho)_t$ a.s., see Lemma (\ref{conv:est}). In terms of $(J_{h,t}^{(n)})_t$, the rejection ellipse $\hat{\mathscr E}:=\hat\Gamma^{1/2}\cdot\mathscr C$ is time dependent.
Under $\mathcal H_0$ estimation is consistent and the asymptotic  $\alpha$-level is kept. 
Third, we propose a conservative approach to avoid estimation. Define a square $\mathscr S := \{(x,y)^\mathsf{T}\in\rn^2\; |\; d_\infty((x,y)^\mathsf{T}) \le Q \}$, while $d_\infty((x,y)^\mathsf{T}):=\max(|x|,|y|)$ denotes the maximum norm. $\mathscr S$ has center zero and edge length equal to the diameter of $\mathscr C$. Because $\Gamma$ has unit diagonals, any ${\mathscr E}$ is trapped in $\mathscr S$, see Figure \ref{fig2}A. Thus, with $\mathscr R:=\mathscr S^c$ the asymptotic probability to falsely reject is less than $\alpha$. The test statistic is
\begin{align}\label{M_infty}
	M_h^{(n)}(\infty) := \max_t d_{\infty}(J_{h,t}^{(n)}).
\end{align}
This approach is optimal in the sense that $\mathscr E$ touches each edge of $\mathscr S$ exactly once, thus $\mathscr S$ can not be shrunk. 
The tests are summarized as follows, see also Table  \ref{three_choices}. 
\begin{theo}\label{theo_main}
	Let $\mathbf X\in\mathcal M$, and for $\alpha \in (0,1)$ let $Q$ be the $(1-\alpha)$-quantile of the limit in (\ref{Mh}).	
	Assume $\mathcal H_0: C=\emptyset$. Then it holds $\lim_{n\to\infty}\pr(M_h^{(n)}(\Gamma)>Q)=\alpha$.
	Further, it holds first $\lim_{n\to\infty}\pr(M_h^{(n)}(I)>Q)=\alpha$ given $\rho=0$,  second $\lim_{n\to\infty}\pr(M_h^{(n)}(\hat \Gamma)>Q)=\alpha$, and third $\lim_{n\to\infty}\pr(M_h^{(n)}(\infty)>Q)\le \alpha$.
\end{theo}


\begin{table}[h!]	
	\centering	
	\begin{footnotesize}
		\begin{tabular}{ | c | c | l | c |}
			\hline
			test statistic & asymp.~level &rejection area $\mathscr R$ w.r.t.~$(J_{h,t})_t$ & assumption \\ 
			\hline
			\hline
			$M_h(I)$ \hfill (\ref{Mh}) & $=\alpha$ &$\mathscr C^c\;=\{(x,y)^\mathsf{T}\in\rn^2\; |\; d_I((x,y)^\mathsf{T}) > Q \}$  & symmetry  \\ \hline
			$M_h(\hat\Gamma)$ \hfill (\ref{M_gamma}) & $=\alpha$ &$\hat{\mathscr E}^c = \hat \Gamma \cdot \mathscr C^c$, time dependent  & / \\ \hline
			$M_h(\infty)$ \hfill (\ref{M_infty}) & $<\alpha$ &$\mathscr S^c= \{(x,y)^\mathsf{T}\in\rn^2\; |\; d_\infty((x,y)^\mathsf{T}) > Q \}$ & / \\
			\hline
		\end{tabular}
	\end{footnotesize}
	\caption{Three ways of testing.}
	\label{three_choices}
\end{table}

Note that Proposition \ref{prop:jhc} below yields asymptotic power one. Further, the bivariate approach avoids double testing unlike the two univariate approaches from Section \ref{sect:idea}: Set $\alpha=5\%$. With the univariate approaches we obtain the boundaries $\approx\pm 3.59$, by simulation of the quantile of the temporal maximum of a single component of $(\mathcal L_{h,t})_t$, see Figure \ref{fig1}B and C. With the bivariate procedure we get $Q\approx 4.00$ (radius in D). This slight increase is the price for avoiding $\alpha$-error accumulation. 

\section{Change point detection}\label{sect:cpd}
\begin{figure}[h!]
	\centering  
	\includegraphics[width=0.5\textwidth,angle=0]{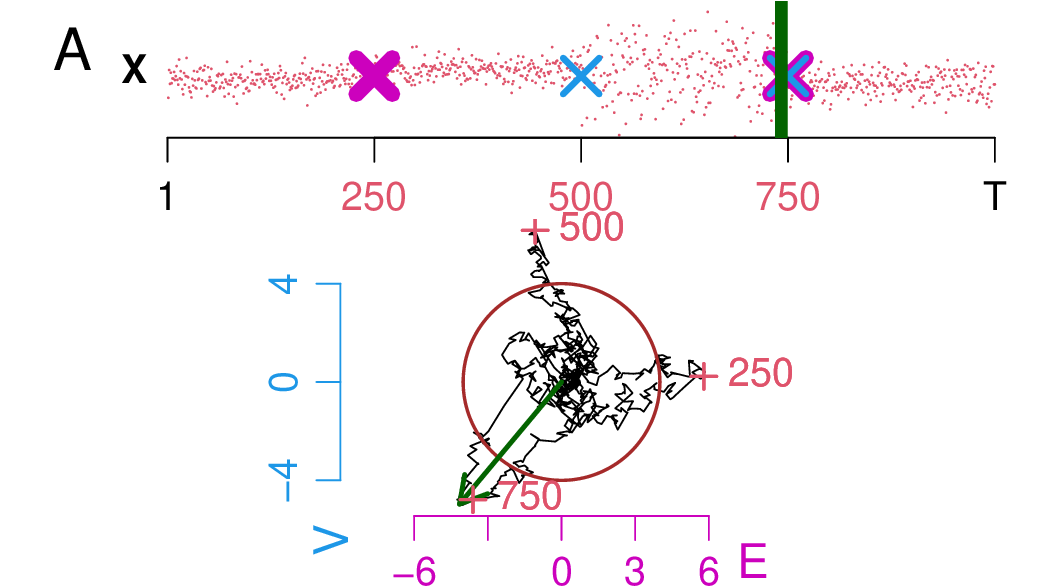}
	\includegraphics[width=0.5\textwidth,angle=0]{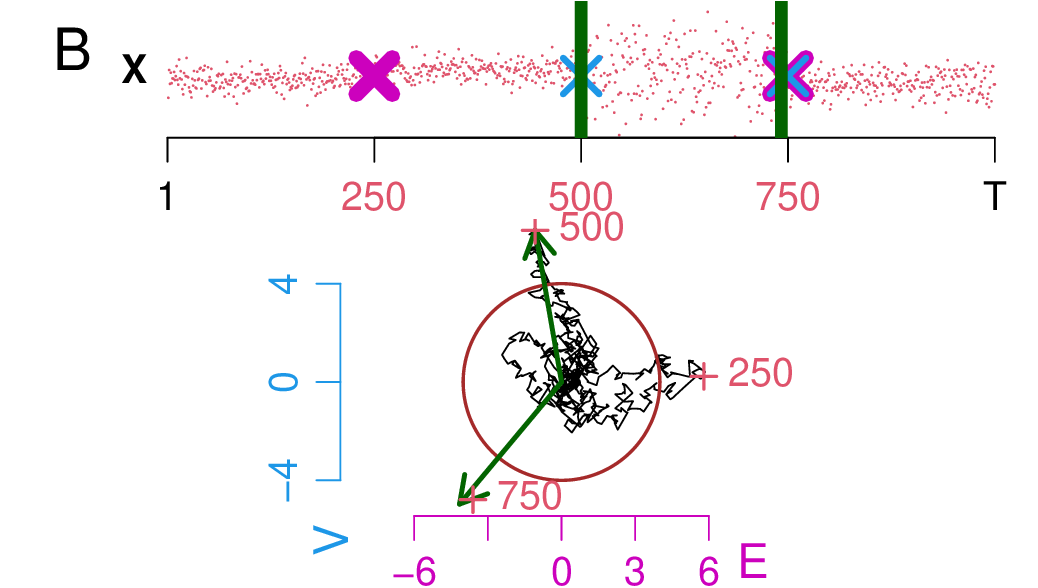}\\
	\vspace{0.5em}
	\includegraphics[width=0.5\textwidth,angle=0]{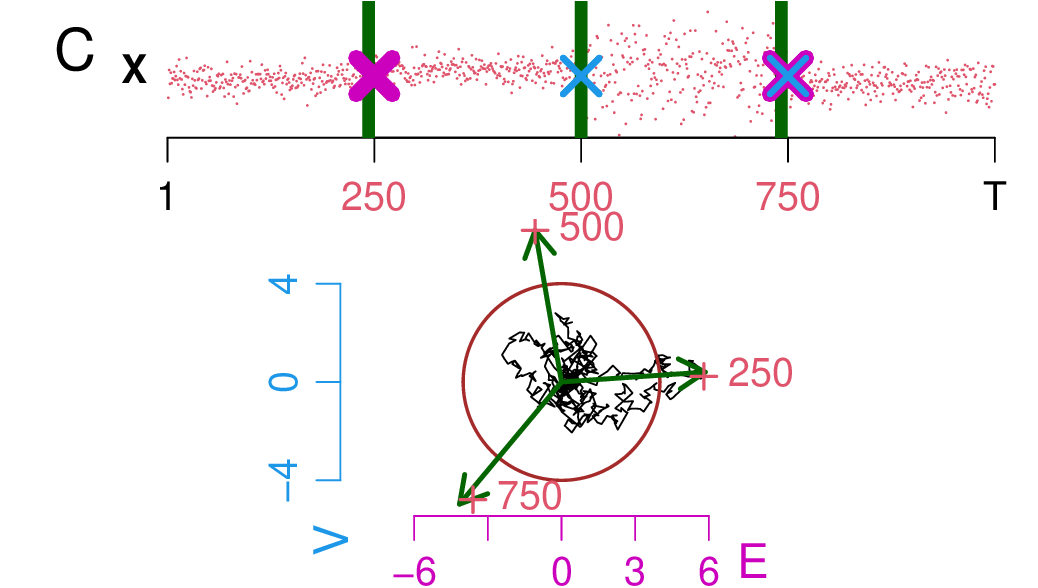}
	\includegraphics[width=0.5\textwidth,angle=0]{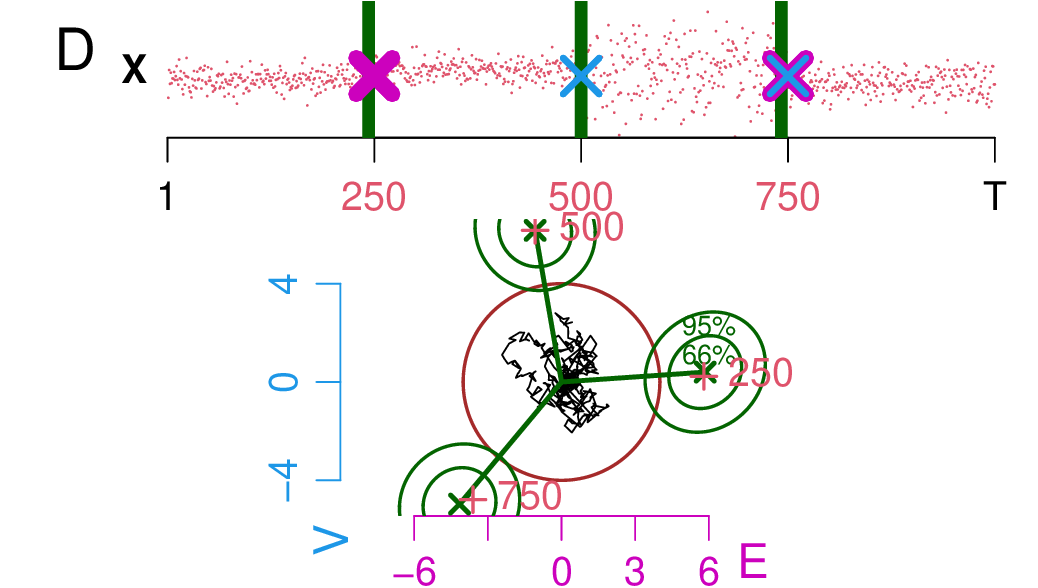}
	\caption{Change point detection via successive $\arg\max$ estimation, setup from Figure \ref{fig1}II. A,B,C: Obtain new estimate $\hat c$ (green bar) as the integer satisfying $J_{h,\hat c}=\max_t d_I(J_{h,t})$ (green arrow). Cut out $(J_{h,t})_t$ in the $h$-neighborhood of $\hat c$ (B,C,D). D: Stop as the remaining $(J_{h,t})_t$ lies in $\mathscr R^c=\mathscr C$ (brown). Estimate the distribution of $J_{h,\hat c}$ (green dartboards).}
	\label{fig3}
\end{figure}

After rejection of $\mathcal H_0$
we detect change points via successive $\arg\max$ estimation, see  Figure \ref{fig3}. Proposition \ref{prop:jhc} below states asymptotic normality of $J_{h,c}^{(n)}$ at $c\in C$, see dartboards in Figures \ref{fig1} and \ref{fig2}. This facilitates inference about the effects, i.e., the strength and the type of the change, and enables interpretation in practice.  

\begin{algo}\label{algo:sfa} Set $\mathscr R$ as either $\mathscr C^c$, $\hat{\mathscr E}^c$ or $\mathscr S^c$ as used in the test. Set  $\hat C_h := \emptyset$ and $\tau_h:=[h,T-h]\cap\nn$.
	While $J_{h,t}\in\mathscr R$ for any $t\in\tau_h$, update $\hat C_h$ and $\tau_h$ as follows: among all $t\in \tau_h$ for which $J_{h,t}\in\mathscr R$, choose the candidate $\hat c$ that maximizes the Euclidean distance of $J_{h,t}$, add $\hat c$ to $\hat C_h$ and delete its $h$-neighborhood $\{\hat c - h+1,\ldots,\hat c + h\}$ from $\tau_h$.
\end{algo}


Note that for any choice of $\mathscr R$, either $\mathscr C^c$, $\mathscr E^c$ or $\mathscr S^c$, it is always the Euclidean distance that inference is based on, see interpretation below. For single $E_{h,t}^{(n)}$ see similar procedures in \citet{Antoch1999} and \citet{Eichinger2018}.

\paragraph{Asymptotic normality of $J_{h,c}^{(n)}$}
\begin{prop}\label{prop:jhc}
	Let $\mathbf X\in\mathcal M$, $h$ be a window size, and $c\in[h,T-h]$ a change point such that $(c-h,c+h]\backslash \{c\}$ is free from change points. Then it holds as $n\to\infty$
	\begin{align*}
		J_{h,c}^{(n)} - \Delta_{h,c}^{(n)} \cdot j_{h,c}^{(n)} 
		\stackrel{d}{\longrightarrow}
		N_2(0,\Gamma_c),
	\end{align*}
	with $\Delta_{h,c}^{(n)}$ and $j_{h,c}^{(n)}$ in (\ref{jhc}) and  $\Gamma_c$ in (\ref{Gammac}).
\end{prop}
Near $c:=c_u$ the process $\mathbf X$ derives from $\mathbf X_u$ and $\mathbf X_{u+1}$. W.l.o.g.~let $u=1$. Set 
\begin{small}
	\begin{align}\label{jhc}
		j_{h,c}^{(n)} :=
		\begin{pmatrix}
			\dfrac{\mu_2-\mu_1}{[(\sigma_2^2 + \sigma_1^2)/(nh)]^{1/2}}\\
			\dfrac{\sigma_2^2-\sigma_1^2}{[(\nu_2^2 + \nu_1^2)/(nh)]^{1/2}}\\
		\end{pmatrix}
		\quad\!\!\textrm{and}\!\!\quad
		\Delta_{h,c}^{(n)} :=
		\begin{pmatrix}
			\left(\dfrac{\sigma_2^2 + \sigma_1^2}{\hat\sigma_r^2 + \hat\sigma_\ell^2}\right)^{1/2} & 0 \\
			0 & \left(\dfrac{\nu_2^2 + \nu_1^2}{\hat\nu_r^2 + \hat\nu_\ell^2}\right)^{1/2}\\
		\end{pmatrix},
	\end{align}
\end{small}
and call $j_{h,c}^{(n)}$  the asymptotic expectation of $J_{h,c}^{(n)}$. 
Further, $\Delta_{h,c}^{(n)}$ replaces the true scaling of $J_{h,c}^{(n)}$ with the local estimators. Lemma \ref{conv:est} implies $\Delta_{h,c}^{(n)}\to I$ a.s.~componentwise as $n\to\infty$. The asymptotic correlation matrix is
\begin{align}\label{Gammac}
	\Gamma_c := 
	\begin{pmatrix}
		1 & \rho_c\\
		\rho_c & 1
	\end{pmatrix}
	\qquad\textrm{with}\qquad
	\rho_c := \frac{\mu_2^{\{3\}} + \mu_1^{\{3\}} }{(\sigma_2^2+\sigma_1^2)^{1/2}\cdot(\nu_2^2 + \nu_1^2)^{1/2}}.
\end{align}
Note the analogy of $\rho_c$ and its estimator $\hat\rho$ in (\ref{hat_rho}). In Figures \ref{fig1} and \ref{fig2} the distribution of $J_{h,c}$ is approximated by $N_2(j_{h,c}^{(1)},\Gamma_c)$ for the three change points. The golden dartboards describe the $66\%$- and $95\%$-contour lines with center  $j_{h,c}^{(1)}$. A stronger $|\rho_c|$ increases the squeezing of the dartboards in Figure \ref{fig2}B. Symmetry $\rho_c=0$ implies circles in Figure \ref{fig1}.
The size of a dartboard is unique up to squeezing as $\Gamma_{c}$ has unit diagonals: $N_2(j_{h,c}^{(1)},\Gamma_c)$ implies that for $\alpha\in(0,1)$ any $\alpha$-contour ellipse is trapped optimally in a square with edge lengths $2\sqrt{q_\alpha}$, while $q_\alpha$ denotes the $\alpha$-quantile of the $\chi^2$-distribution with two degrees of freedom. 

\paragraph{Change point interpretation}
To interpret effects we decompose
$
j_{h,c}^{(n)} = \sqrt{nh} \cdot  d_I(j_c) \cdot 
(\cos(\omega),\sin(\omega))^{\mathsf{T}}
$,
with $j_{c} := j_{h,c}^{(n)}/\sqrt{nh}$, and $\omega\in[0,2\pi)$ the angle between $j_{c}$ and the abscissa. Then $j_c$ is a signal to noise ratio that captures the effects: the Euclidean distance $d_I(j_c)$ measures the strength of the change, and the angle $\omega$ the type of the change, with $\omega \in	\{0,\pi\}  \Leftrightarrow$ change only in expectation $(0 = \textrm{increase}, \pi = \textrm{decrease})$, and $\omega\in	\{\pi/2,(3/2)\pi\}  \Leftrightarrow$ change only in variance $(\pi/2 = \textrm{increase}, (3/2)\pi = \textrm{decrease})$. Generally, a dartboard is shifted towards $(cos(\omega),sin(\omega))^\mathsf{T}$, see Figure \ref{fig1} or \ref{fig2}. A dartboard lies far off of zero if either $\sqrt{nh}$ or $d_I(j_c)$ is large. 
This implies asymptotic power one, also note extensions to local changes where the parameters imply $d_I(j_c)=o(1/\sqrt{nh})$, see e.g., \citet{Antoch1999}.
 Further note that the location does not depend on the symmetry of $\mathbf{X}$ which is captured in $\Gamma_c$. Thus, in Algorithm \ref{algo:sfa} it is always the Euclidean distance that $J_{h,t}^{(n)}$ is judged by, regardless of $\mathscr R$. 

In practice, for an estimate $\hat c$ interpretation is based on $J_{h,\hat c}$, see green arrows in Figure \ref{fig3}.
 We assign the $66\%$- and $95\%$-contour lines of $N_2(J_{h,\hat c},\hat\Gamma_{\hat c})$ with $\hat\Gamma_{\hat c}$ as in (\ref{hat_rho}), see green dartboards in Figure \ref{fig3}D. 
This strengthens confidence in interpretation: it is plausible that the smallest estimate refers to a change only in $\mu$ (right dartboard close to abscissa) and that the middle one indicates a change only in $\sigma^2$ (upper dartboard close to ordinate). A single parameter change at the largest estimate, however, is rather implausible (dartboard bottom left). 


\section{The joint process in case of change points}\label{sect:jht_c}
Let $\mathbf X\in \mathcal M$ with $\Gamma_u=I$ (symmetry). Proposition \ref{prop_main2} below  yields $(J_{h,t})_t \stackrel{d}{\approx} (j_{h,t})_t + (\mathcal L_{h,t}^*)_t$ approximately, with non-random $(j_{h,t})_t$ describing excursions from zero in the $h$-neighborhood of any $c_u$. $(\mathcal L_{h,t}^*)_t$ is a Gaussian process with $\mathcal L_{h,t}^*\sim N_2(0,I)$.


\begin{figure}[h!]
	\centering  
	\hrule\vspace{0.5em}
	\includegraphics[width=0.5\textwidth,angle=0]{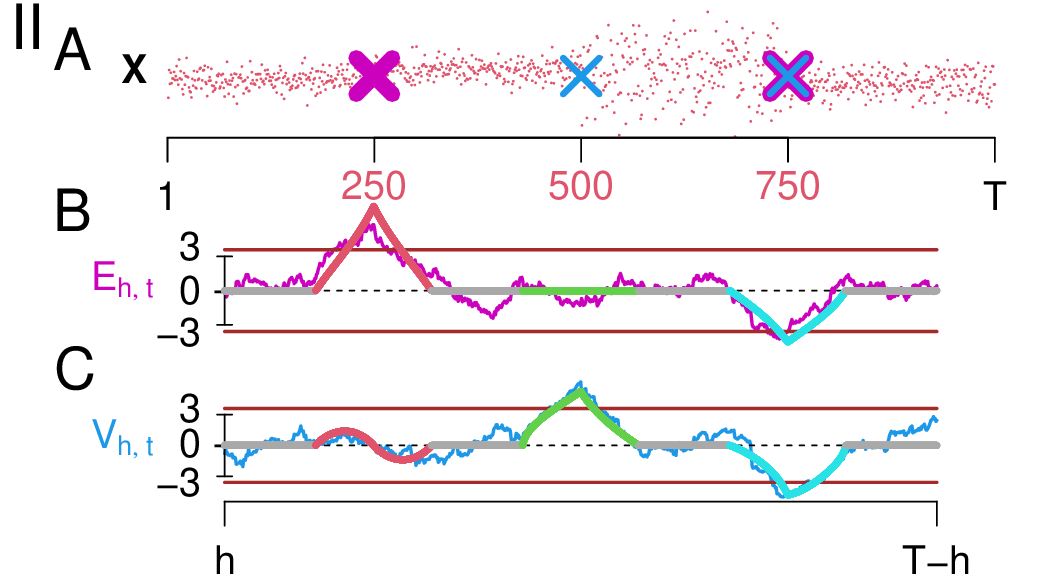}
	\includegraphics[width=0.5\textwidth,angle=0]{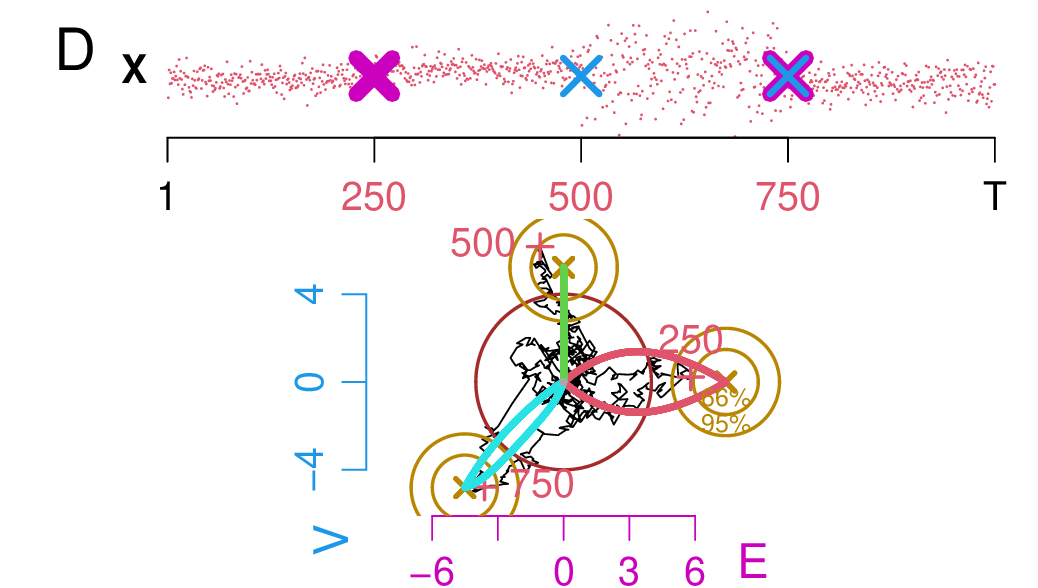}\\
	\vspace{0.5em}\hrule\vspace{0.5em}
	\includegraphics[width=0.5\textwidth,angle=0]{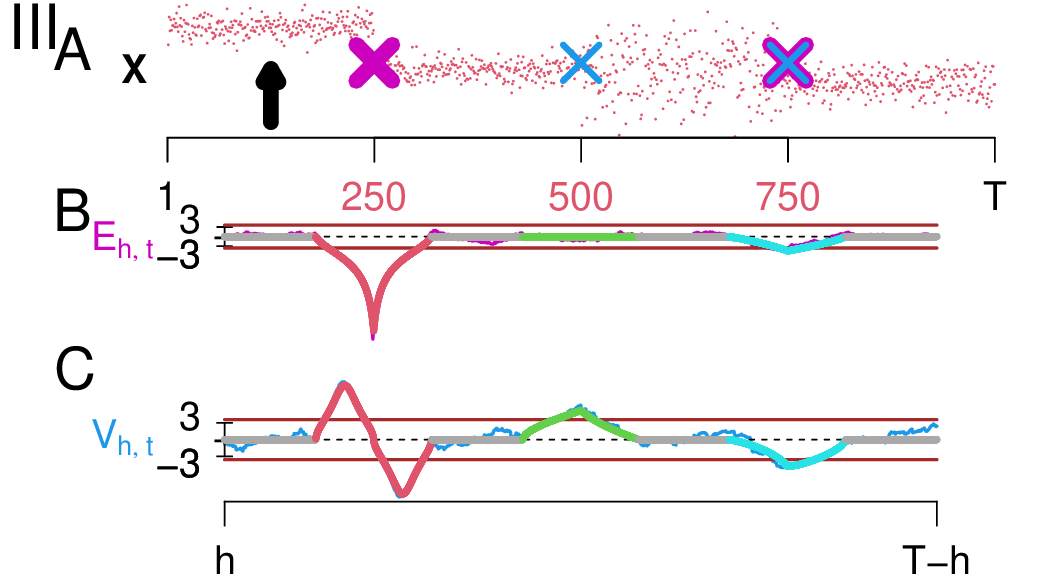}
	\includegraphics[width=0.5\textwidth,angle=0]{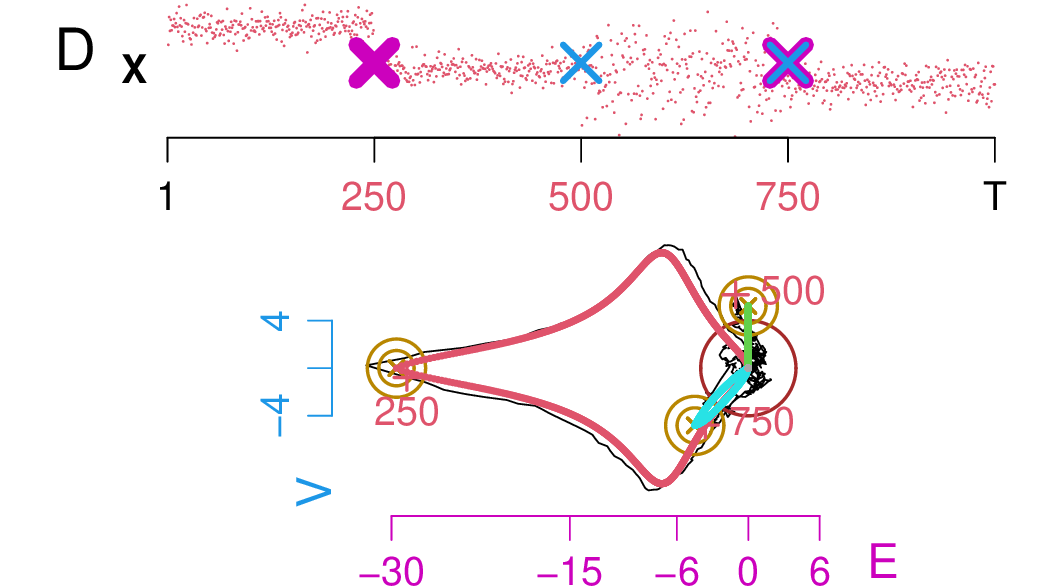}
	\caption{The centering terms $(e_{h,t})_t$ (B), $(v_{h,t})_t$ (C) and $(j_{h,t})_t$ (D), colored inside the $h$-neighborhood of any $c_u$ (red $c_1=250$, green $c_2=500$,  light blue $c_3=750$), else gray at zero. Extension of Figure \ref{fig1} panels II and III.}
	\label{figshark}
\end{figure}

Figure \ref{figshark} extends Figure \ref{fig1}II and III:
$J_{h,t}=(E_{h,t},V_{h,t})^\mathsf{T}$ fluctuates close to $j_{h,t} :=(e_{h,t},v_{h,t})^\mathsf{T}$ (red, green or light blue). 
The sensitivity of $J_{h,t}$ to the effects is explicated in $j_{h,t}$: $e_{h,t}$ is sensitive to changes in $\mu$ and robust against higher order changes, and $v_{h,t}$ reacts to changes in $\sigma^2$ but not to higher order changes. This supports Algorithm \ref{algo:sfa}. Also, $v_{h,t}$ reveals the error caused by a change in $\mu$, see IIIC (red). In IIID (red) the bivariate approach overcomes the error: for any excursion, $d_I(j_{h,t})$ is maximal at $c_u$. We consider $C=\{c\}$ but note direct generalization to any $C$ with $\min_{u} |c_{u}-c_{u-1}|\ge2h$.

\begin{figure}[h!]
	\centering  
	\includegraphics[width=0.65\textwidth,angle=0]{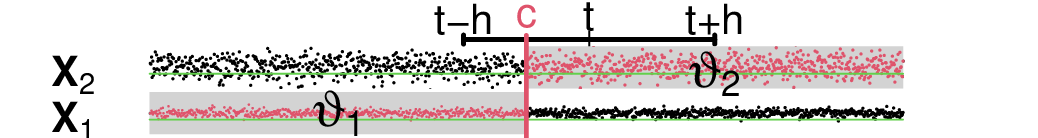}
	\caption{Decomposition for $C=\{c\}$. The left window $(t-h,t]$ decomposes into $(t-h,c]$ referring to $\bold{X}_1$  and $(c,t]$ referring to $\bold{X}_2$. The right window $(t,t+h]$ entirely refers to $\bold{X}_2$.
	}
	\label{fig18}
\end{figure}

\begin{prop}\label{prop_main2}
	Let $\mathbf X\in\mathcal M$ with 
	$C=\{c\}$ and $h$ be a window size such that  $c\in[h,T-h]$, and let $\rho_1=\rho_2=0$. In $(\mathcal D_{\rn^2}[h,T-h],d_{SK})$ it holds as $n\to\infty$
	\begin{align}\label{conv:jht}
	\left[
	\tilde D_{h,t} \cdot
	\left(J_{h,t}^{(n)} - \Delta_{h,t}^{(n)}\cdot j_{h,t}^{(n)}\right)
	\right]_t
	\stackrel{d}{\longrightarrow} (\mathcal L_{h,t}^*)_t.
	\end{align}
\end{prop}
$(\mathcal L_{h,t}^*)_t$ is a bivariate continuous $2h$-dependent process with $\mathcal L_{h,t}^*\sim N_2(0,I)$ for all $t$, and for $t\not\in (c-h,c+h)$ it equals $(\mathcal L_{h,t})_t$ from (\ref{limit}). We give $(j_{h,t}^{(n)})_t$ in (\ref{eandv}), and $(\tilde D_{h,t})_t$ and  $(\Delta_{h,t}^{(n)})_t$ in (\ref{remaining}). The result extends Proposition \ref{prop_main}. The marginal at $c$ is Proposition \ref{prop:jhc}. 
We first extend Lemma \ref{conv:est}.
\begin{lemm}\label{lemm:conv_est_c}
	Let $\mathbf X\in\mathcal M$ with 
	$C=\{c\}$ and $h$ a window size such that  $c\in[h,T-h]$.
	For $j\in\{\ell,r\}$ it holds in $(\mathcal D_{\rn}[h,T-h],d_{\|\cdot\|})$ as $n\to\infty$ almost surely $(\hat\mu_j)_t\to(\tilde\mu_j)_t$,	$(\hat\sigma_j^2)_t\to (\tilde\sigma_j^2)_t$ and $(\hat\nu_j^2)_t\to (\tilde\nu_j^2)_t$.
\end{lemm}	
See (\ref{tilde_vartheta}) and (\ref{tilde_left}) for the limits. They are non-random functions in $t$, in $(c-h,c+h)$ depending on parameters of both $\mathbf X_1$ and $\mathbf{X}_2$.
The tilde marks dependence on $t$. For $j\in\{\ell,r\}$ set $(\tilde\vartheta_j)_t$ as a placeholder for the limits, and $\vartheta_1$ and $\vartheta_2$ as the associated parameters. E.g., choose $\tilde\mu_\ell$ with $\mu_1$ and $\mu_2$, see Figure \ref{fig18}.


If a window does not lap $c$, then the RVs refer to a single $\mathbf X_u$ and $\tilde \vartheta_j$ equals $\vartheta_u$,
\begin{align}\label{tilde_vartheta}
\tilde\vartheta_\ell &:= 
\begin{cases}
\vartheta_1,& t<c,\\
\vartheta_2,& t\ge c+h,
\end{cases} 
\quad\qquad\textrm{and}\quad\qquad
\tilde\vartheta_r := 
\begin{cases}
\vartheta_1,& t<c-h,\\
\vartheta_2,& t\ge c,
\end{cases}	
\end{align}
e.g., $\tilde\vartheta_r=\vartheta_2$ in Figure \ref{fig18}, also see Lemma \ref{conv:est}. If a window laps $c$, then $\tilde\vartheta_j$ depends on both $\mathbf X_1$ and $\mathbf X_2$, see left window in Figure \ref{fig18}. Set $\tilde\vartheta_\ell$ for $t\in [c,c+h)$
	\begin{align}\label{tilde_left}
	\tilde\mu_\ell & := \frac{c-(t-h)}{h} \mu_1 \;+\; \frac{t-c}{h} \mu_2\\
	\tilde\sigma_\ell^2 & := \frac{c-(t-h)}{h} [\sigma_1^2 +(\tilde\mu_\ell -\mu_1)^2] \;+\; \frac{t-c}{h} [\sigma_2^2+(\tilde\mu_\ell -\mu_2)^2]\nonumber\\
	\tilde\nu_\ell^2 & :=
	\Big(\frac{c-(t-h)}{h} [\mu_1^{\{4\}} - 4\mu_1^{\{3\}}(\tilde\mu_\ell-\mu_1) + 6\sigma_1^2(\tilde\mu_\ell-\mu_1)^2 + (\tilde\mu_\ell-\mu_1)^4]\nonumber\\
	&\qquad\qquad + \frac{t-c}{h} [\mu_2^{\{4\}} - 4\mu_2^{\{3\}}(\tilde\mu_\ell-\mu_2) + 6\sigma_2^2(\tilde\mu_\ell-\mu_2)^2 + (\tilde\mu_\ell-\mu_2)^4] \Big) - \tilde\sigma_\ell^4,\nonumber
	\end{align}
and define $\tilde\vartheta_r$ for $t\in[c-h,c)$ by replacing in (\ref{tilde_left}) all subscripts $\ell$ with $r$ and $t$ with $t+h$. We comment on the left window  $(t-h,t]$. A proportion of $[c-(t-h)]/h$ RVs belongs to $\mathbf{X}_1$ and $(t-c)/h$ to $\mathbf{X}_2$. Then $\tilde\mu_\ell$ is a linear interpolation between $\mu_1$ and $\mu_2$. The higher orders are further affected by $\tilde\mu_\ell$ via the error $(\tilde\mu_\ell-\mu_u)^k$ with $k\in\{1,2,4\}$. E.g., in $\tilde\sigma_\ell^2$ we find $(\tilde\mu_\ell-\mu_u)^2$ deriving from $\hat\sigma_\ell^2 = (1/nh)\sum_{i\in I_\ell}(X_i - \hat \mu_\ell)^2$ as $\hat \mu_\ell$ relies on all RVs in $(t-h,t]$ irrespective of $c$. All errors vanish iff  $\mu_1=\mu_2=\tilde\mu_\ell$. 

We now define the centering $j_{h,t}^{(n)} =(e_{h,t}^{(n)},v_{h,t}^{(n)})^\mathsf{T}$ from $(\ref{tilde_vartheta})$ and $(\ref{tilde_left})$ via 
\begin{align}\label{eandv}
e_{h,t}^{(n)} := \frac{\tilde\mu_r- \tilde\mu_\ell}{[(\tilde\sigma_r^2 + \tilde\sigma_\ell^2)/(nh)]^{1/2}} 
\qquad\textrm{and}\qquad
v_{h,t}^{(n)} := \frac{\tilde\sigma_r^2- \tilde\sigma_\ell^2}{[(\tilde\nu_r^2 + \tilde\nu_\ell^2)/(nh)]^{1/2}},
\end{align}
see $E_{h,t}^{(n)}$ and $V_{h,t}^{(n)}$ in (\ref{EandV}). For $t\in (c-h,c+h)$ we discuss $e_{h,t}$ and $v_{h,t}$ w.r.t.~the type of change. For $e_{h,t}$, see Figure \ref{figshark}B. Message 1: constant expectation implies $e_{h,t}=0$ (green). This is because $\mu_1=\mu_2$ implies $\tilde\mu_{\ell}=\tilde\mu_r$ see (\ref{tilde_left}), regardless of higher moments. Thus, $E_{h,t}^{(n)}$ is robust against higher order changes. 
Message 2: a change in expectation causes $e_{h,t}$ to deviate from zero (red or light blue). Thus, $E_{h,t}^{(n)}$ is sensitive to expectation changes. $\tilde\mu_j$ linearly interpolates $\mu_1$ and $\mu_2$, thus  $\tilde\mu_r - \tilde\mu_{\ell}$ has a hat shape peaking at $c$ and is additionally scaled by $(\tilde\sigma_r^2 + \tilde\sigma_{\ell}^2)^{1/2}$.

For $v_{h,t}$ see Figure \ref{figshark}C. Complexity rises as a change in expectation affects higher moments. Message 1: constant expectation results in $v_{h,t}$ behaving analogously to $e_{h,t}$: first, constant variance implies $v_{h,t}=0$, as for $\mu_1=\mu_2$ and $\sigma_1^2=\sigma_2^2$ we find $\tilde\sigma_{\ell}^2=\tilde\sigma_r^2$, see (\ref{tilde_left}). Second, a change in variance causes a deviation from zero (green): if $\mu_1=\mu_2$, then $\tilde\sigma_j^2$ linearly interpolates $\sigma_1^2$ and $\sigma_2^2$, thus $\tilde\sigma_r^2-\tilde\sigma_{\ell}^2$ has a hat shape that is scaled with $(\tilde\nu_r^2 + \tilde\nu_{\ell}^2)^{1/2}$. Thus, given constant expectation, $V_{h,t}^{(n)}$ is sensitive to changes in variance and robust against higher order changes. Regarding $j_{h,t}$ note a linear trajectory along the ordinate, see Figure \ref{figshark}D (green).
Message 2: a change in expectation falsely results in a systematic deviation of $v_{h,t}$ from zero even if the variance is constant (red).
Thus, $V_{h,t}^{(n)}$ is also sensitive to changes in expectation.

We state the remaining functions of (\ref{conv:jht}). 
We set $(\Delta_{h,t}^{(n)})_t$ and $(\tilde D_{h,t})_t$ as
\begin{align}\label{remaining}
\Delta_{h,t}^{(n)} :=
\begin{pmatrix}
\Delta_{h,t,1}^{(n)} & 0 \\
0 & \Delta_{h,t,2}^{(n)}\\
\end{pmatrix}
\qquad\textrm{and}\qquad
\tilde D_{h,t} := 
\begin{pmatrix}
\tilde D_{h,t,1} & 0 \\
0 & \tilde D_{h,t,2}\\
\end{pmatrix},
\end{align}
with 
$\Delta_{h,t,1}^{(n)}:=[(\tilde\sigma_r^2 + \tilde\sigma_\ell^2)/(\hat\sigma_r^2 + \hat\sigma_\ell^2)]^{1/2}$ and 
$\Delta_{h,t,2}^{(n)}:=[(\tilde\nu_r^2 + \tilde\nu_\ell^2)/(\hat\nu_r^2 + \hat\nu_\ell^2)]^{1/2}$ as well as
$\tilde D_{h,t,1}:=\lim_{n\to\infty}[(\tilde\sigma_r^2 + \tilde\sigma_\ell^2)/(nh \va(\hat\mu_r - \hat\mu_\ell))]^{1/2}$ and  $\tilde D_{h,t,2}:=\lim_{n\to\infty}[(\tilde\nu_r^2 + \tilde\nu_\ell^2)/(nh \va(\hat\sigma_r^2 - \hat\sigma_\ell^2)) ]^{1/2}$. 

$\Delta_{h,t}^{(n)}$ replaces the limits with the estimators. It is $(\Delta_{h,t}^{(n)})_t\to(I)_t$ a.s.~componentwise as $n\to\infty$, see Lemma \ref{lemm:conv_est_c}. $\Delta_{h,c}^{(n)}$ is (\ref{jhc}). $\tilde D_{h,t}$ replaces the limits with the true scaling of the numerators of $J_{h,t}^{(n)}$ yielding asymptotic unit variance. $\tilde D_{h,t}$ captures inconsistent parameter estimation. Outside $(c-h,c+h)$ and at $c$ both windows in their entirety refer to a single $\mathbf X_u$, stating consistency $\tilde D_{h,t} = I$. In Figure \ref{figshark}IIID (red) $J_{h,t}$ is close to $j_{h,t}$ but slightly shifted, as we had to distort $j_{h,t}$ with $\tilde D_{h,t}$. Note that $nh\va(\hat\mu_\ell)$ considers $c$ while $\hat\sigma_\ell^2$ does not: if $(t-h,t]\not\ni c$ then $nh\va(\hat\mu_\ell)\to\sigma_u^2$ as $n\to\infty$. If $(t-h,t]\ni c$ then $nh\va(\hat\mu_\ell)\to [(c-(t-h))/h]\sigma_1^2 +[(t-c)/h]\sigma_2^2$, note the absence of the error $(\tilde\mu_\ell -\mu_1)^2$ in contrast to $\tilde\sigma_\ell^2$ in (\ref{tilde_left}). 


\section{Practical Performance}\label{sect:practical}
In Subsection \ref{subsect:multiwindow} we extend methodology to multiple windows. This improves the detection of change points on multiple time scales and with different effects. In Subsection \ref{subsect:R} we discuss the companion \texttt{R}-package $\texttt{jcp}$. In Subsection \ref{subsect:level_window} we provide simulation studies for $\mathcal H_0: C=\emptyset$. We find the asymptotic significance level $\alpha$ of the test to be kept in many scenarios. In Subsection \ref{subsect:performance} we discuss simulations for $C\not=\emptyset$. We confirm reliable detection accuracy and adequate interpretation. To conclude, we present a real data example in Subsection \ref{subsect:data}.

\subsection{Extension to multiple windows}\label{subsect:multiwindow}
For $\mathbf X\in\mathcal M$ let $H$ be a set of increasingly ordered windows $h_1<\cdots <h_w$, yielding multiple processes $\{(J_{h,t})_t \;|\; h\in H\}$. Smaller $h_k$ are sensitive to rapid changes, as they do no overlap adjacent change points and thus ensure an unbiased excursion of $(J_{h,t})_t$ which supports precise detection. Larger $h_k$ improve detection of small effects, as at a $c\in C$ a dartboard is shifted outwards with order $\sqrt{nh}$, see Proposition \ref{prop:jhc}. 

We extend the test: for $(\mathcal D_{\rn^2}[h_k,T-h_k],d_{SK})$, $k=1,\ldots,w$ let 
 $(\times_{k=1}^{w}\mathcal D_{\rn^2}[h_k,T-h_k],d_{SK}^w)$ denote the product space with distance $d_{SK}^w$ metrizing the product topology. 
Poposition \ref{prop_main} extends to

\begin{coro}\label{coro_main}
	Let $\mathbf X\in\mathcal M$ with $C=\emptyset$ and $H=\{h_1,\ldots,h_w\}$ be a set of windows. 
	 In $(\times_{k=1}^{w}\mathcal D_{\rn^2}[h_k,T-h_k],d_{SK}^w)$ it holds as $n\to\infty$
	\begin{align*}
	(\Gamma^{-1/2}\cdot J_{h_1,t}^{(n)},\ldots,\Gamma^{-1/2}\cdot J_{h_w,t}^{(n)})_t
	\stackrel{d}{\longrightarrow}
	(\mathcal L_{h_1,t},\ldots,\mathcal L_{h_w,t})_t,
	\end{align*}
	with $\Gamma$ in (\ref{gamma}), and $(\mathcal L_{h_k,t})_t$ as in (\ref{limit}) being  evaluated on a single planar Brownian motion over all $k=1,\ldots,w$.
\end{coro}	
Under $C=\emptyset$ continuous mapping yields convergence of the global maximum
\begin{align}\label{M}
	M^{(n)}(\Gamma) := \max_{h\in H} M_h^{(n)}(\Gamma)
	=\max_{h\in H}\max_{t\in[h,T-h]} d_\Gamma(J_{h,t}^{(n)})
	\stackrel{d}{\longrightarrow}
	\max_{h\in H} \max_{t\in [h,T-h]} d_I(\mathcal L_{h,t}),
\end{align}
for $n\to\infty$, which extends (\ref{Mh}). 
We use  $M(\Gamma):=M^{(1)}(\Gamma)$ as a test statistic and reject iff it exceeds 
the $(1-\alpha)$-quantile $Q$ of the limit distribution. The latter is approximated in simulations. We treat $\Gamma$ analogously to the case of single $h$, see Table \ref{three_choices}: first, symmetry assumption yields $\Gamma=I$, and thus $d_I$. Second, consistent estimation $\hat \Gamma$ depends on $t$ and now also on $h$, see (\ref{hat_rho}), yielding $d_{\hat\Gamma}$. Third, using $d_{\infty}$ avoids estimation and is conservative. For the process perspective choose $\mathscr R=\mathscr R(H)$ such that any $(J_{h,t})_{t}$ enters $\mathscr R$ only with probability $\alpha$, if $C=\emptyset$. Let $\mathscr C$ be the circle with center zero and radius $Q$. For the three cases we obtain first $\mathscr R=\mathscr C^c$,  second $\mathscr R =\hat{\mathscr E}^c$ with $\hat{\mathscr E}=\hat \Gamma^{1/2} \cdot \mathscr C$ depending on $t$ and $h$, and third $\mathscr R=\mathscr S^c$ with $\mathscr S$ being the square with center zero and edge length $2Q$.
We extend Theorem \ref{theo_main}.
\begin{theo}\label{theo_main2}
	Let $\mathbf X\in\mathcal M$, and for $\alpha \in (0,1)$ let $Q$ be the $(1-\alpha)$-quantile of the limit in (\ref{M}).	
	Assume 
	$\mathcal H_0: C=\emptyset$. Then it holds $\lim_{n\to\infty}\pr(M^{(n)}(\Gamma)>Q)=\alpha$.
	Further, it holds first $\lim_{n\to\infty}\pr(M^{(n)}(I)>Q)=\alpha$ given $\rho=0$,  second $\lim_{n\to\infty}\pr(M^{(n)}(\hat \Gamma)>Q)=\alpha$, and third $\lim_{n\to\infty}\pr(M^{(n)}(\infty)>Q)\le \alpha$.
\end{theo}

If $\mathcal H_0$ is rejected we aim at estimating $C$. We use a bottom-up approach similar to \citet{Messer2014}. First detect candidates $\hat C_{h_k}$ for each $h_k$.  Then merge the candidates into a final set $\hat C$ favoring smaller over larger windows. 

\begin{algo}\label{algo:mfa} Set $\mathscr R(H)$ as either $\mathscr C^c$, $\hat{\mathscr E}^c$ or $\mathscr S^c$ as used in the test. For each $h_k\in H$ obtain $\hat C_{h_k}$ using Algorithm \ref{algo:sfa} w.r.t.~$\mathscr R(H)$. Set $\hat C :=\hat C_{h_1}$. For increasing $k=2,3,\ldots w$ update $\hat C$ as follows: add any $\hat c \in \hat C_{h_k}$ to $\hat C$ that satisfies $\{\hat c - h_k + 1, \ldots, \hat c + h_k\}\,\cap\,\hat C \,=\, \emptyset$.
\end{algo}


\begin{figure}[h!]
	\centering  
	\includegraphics[width=0.65\textwidth,angle=0]{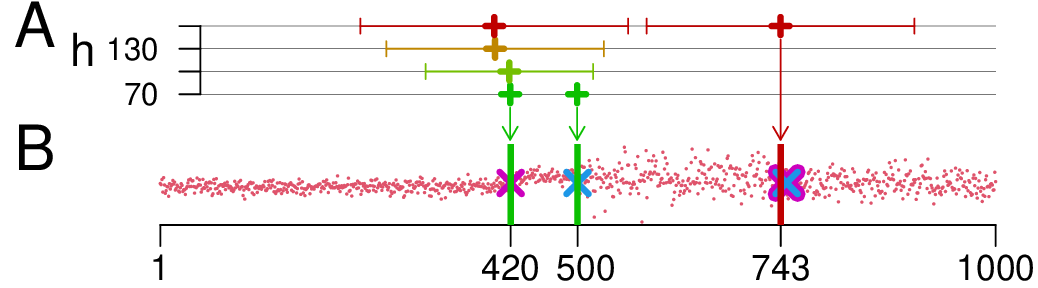}
	\caption{The multi window algorithm.
		B: $\mathbf X$ piecewise i.i.d.~$N(\mu,\sigma^2)$ distributed, $T=1000$, $C=\{420,500,750\}$ (crosses), $\mu=2,10,10,4$ and $\sigma=4,4,12,8$.
		$\mathscr R(H)=\mathscr C^c$, $\alpha=5\%$. Final estimates $\hat C=\{420,500,743\}$ (vertical bars, color-coded according to the window associated with acceptance in A, twice light green and once dark red).  A: $H=\{70,100,130,160\}$ (light green to dark red), candidates $\hat C_{h_k}$ ('+'), their $h_k$-neighborhoods (colored horizontal lines). Algorithm: first set $\hat C:=\hat C_1 =\{420,500\}$, next dismiss $\hat C_2 =\{419\}$ and $\hat C_3 =\{402\}$, finally from $\hat C_4 =\{401,743\}$ only accept $743$.}
	\label{fig19}
\end{figure}

A candidate from the $k$-th step is dismissed if a previously accepted estimate falls into the candidates $h_k$-neighborhood. See Figure \ref{fig19} with $C=\{c_1,c_2,c_3\}$ and $|H|=4$ yielding three estimates $|\hat C|=3$. Note that $|\hat C_k|\le 2$, i.e., no single $h_k$ estimates all three change points, as small $h_k$ lack power to detect the small effect at $c_3$, while larger $h_k$ are not sensitive to the close proximity of $c_1$ and $c_2$.

\subsection{The \texttt{jcp}-package}\label{subsect:R}
\begin{figure}[h!]
	\centering  
	\includegraphics[width=0.6\textwidth,angle=0]{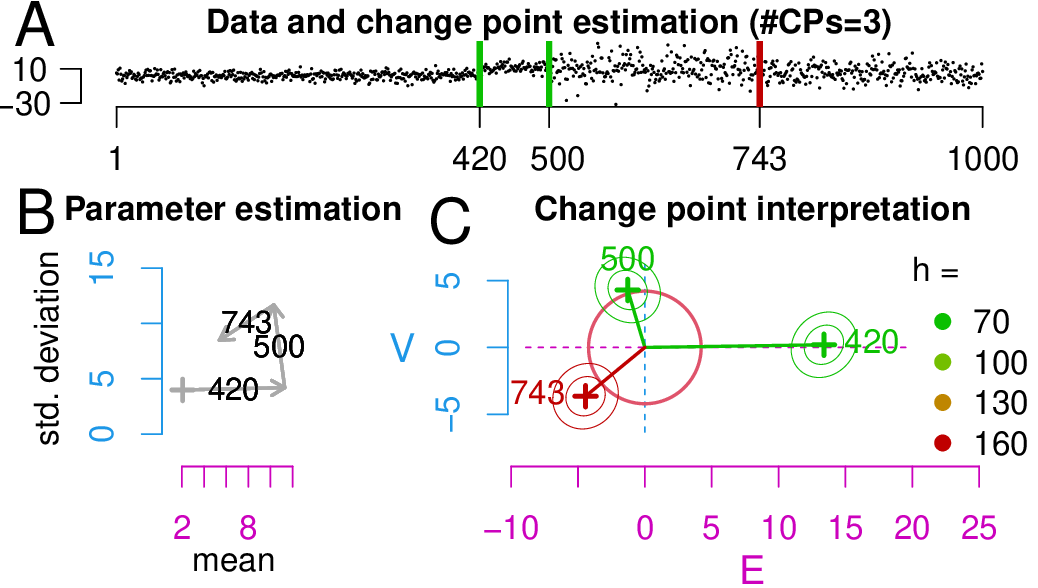}
	\caption{Plotting routine of the \texttt{R}-package \texttt{jcp}.}
	\label{fig20}
\end{figure}

The method is made available in the \texttt{R}-package \texttt{jcp} (\textit{joint change point} detection) on CRAN. The package also contains a summary and a plotting routine. For the example of Figure \ref{fig19} it generates Figure \ref{fig20}.


Panel A shows $\mathbf X$. $C=\emptyset$ is rejected. The bars mark $\hat C=\{420,500,743\}$. Panel B shows the means ($\hat\mu\approx 2.1, 11.3, 10.3, 5.4$) and standard deviations ($\hat\sigma\approx 4.0, 4.2,  11.7, 8.5$) calculated from all $(X_i)_i$ in the estimated sections: we start at '$+$' (first section) and follow the arrows, i.e., point right (increase $\hat\mu$ at $420$), then upwards (increase in $\hat\sigma$ at $500$), and then south-east (decrease in both $\hat\mu$ and $\hat\sigma$ at $743$). The length of an arrow marks the effect size. Panel C facilitates interpretation, recall Section \ref{sect:cpd}: the symmetry assumption yields $\mathscr R=\mathscr C^c$. The $(J_{h,t})_t$ are omitted for overview. The $h_k$ are color-coded (legend). The effect at $\hat c_1=420$ is strong as it is detected by the smallest $h_1=70$ (light green), also the dartboard lies far off. The dartboard touches the abscissa suggesting a change only in $\mu$. The effect at $\hat c_2=500$ is less strong. However, it was still $h_1$ that registered the change point. The dartboard lies close to the ordinate suggesting a change only in $\sigma^2$. The largest $h_4=160$ (dark red) was necessary to detect $\hat c_3=743$. The effect is the smallest among all (compare short arrow in B). The dartboard is positioned diagonally suggesting changes in both parameters. 

\subsection{Significance level and window choice}\label{subsect:level_window}

\paragraph{Significance level}
We constructed an asymptotic test, where $\mathscr R \in\{\mathscr C^c, \hat{\mathscr E}^c\}$ asymptotically keeps the $\alpha$-level, while $\mathscr R=\mathscr S^c$ reduces it, see Table \ref{three_choices} above. We fixed $h$, linked it to $n$ via $nh$, and let $n\to\infty$. In practice however, we set $n=1$. Thus, for a suitable approximation of $Q$ (resp.~$\mathscr R$), we need the smallest window $h_1$ to be sufficiently large. Here we evaluate the rejection probability under $C=\emptyset$ for $n=1$ in simulations. It turns out that $h_1\approx 50$ typically keeps $\alpha$. 

In the following let $\alpha=5\%$. Processes with $T=1000$ and $C=\emptyset$ (here i.i.d.~RVs) are simulated.  The relative frequency of rejections $f_{\mathscr R}$ ($1000$ simulations) approximates the rejection probability. 


\begin{figure}[h!]
	\centering  
	\includegraphics[width=0.3\textwidth,angle=0]{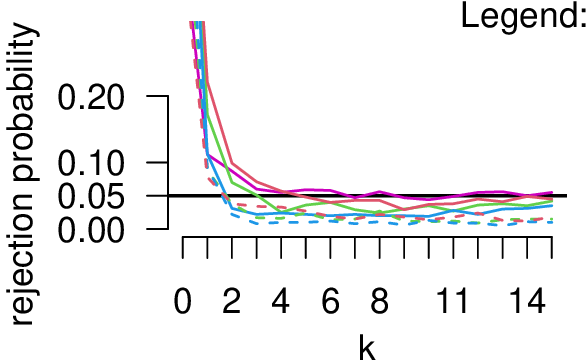}
	\hspace{0em}
	\includegraphics[width=0.3\textwidth,angle=0]{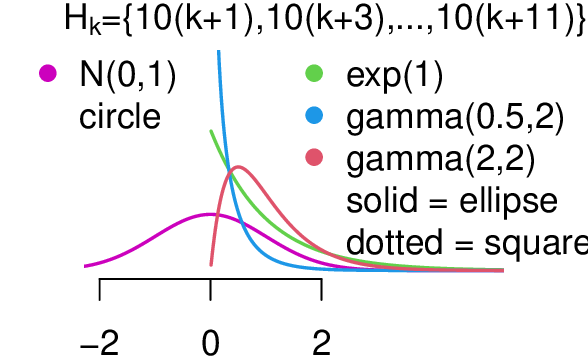}
	\caption{Left: $f_{\mathscr R}$ ($1000$ simulations) for $H_k=\{10(1+k),10(3+k),\ldots,10(11+k)\}$ with $k\in\{0,1,\ldots,15\}$.  $T=1000$, $C=\emptyset$, $\alpha =5\%$. Four distributions color coded: $N(0,1)$ with $\mathscr R=\mathscr C^c$ (magenta). Also $exp(1)$ (green), $gamma(0.5,2)$ (blue) and $gamma(2,2)$ (red) with  $\mathscr R=\hat{\mathscr E}^c$ (solid) or	$\mathscr R=\mathscr S^c$ (dotted). Right: Legend.}
	\label{figrej1}
\end{figure}

Figure \ref{figrej1} shows $f_{\mathscr R}$ as a function of the window set $H_k$, differentiating four distributions: $N(0,1)$ (magenta), $exp(1)$ (green), $gamma(s=1/2,\lambda=2)$ (blue) and
$gamma(2,2)$
(red) with shape $s$ and rate $\lambda$.
For $N(0,1)$ we use $\mathscr R=\mathscr C^c$, else $\mathscr R =\hat{\mathscr E}^c$ (solid lines) or $\mathscr R=\mathscr S^c$ (dotted lines).
We choose $H_k=\{10(1+k),10(3+k),\ldots,10(11+k)\}$ for $k\in\{0,1,\ldots,15\}$, i.e., $|H_k|=6$ and all windows increase by $10$ when $k$ rises one unit. We see that $f_{\mathscr R}$ tends to the true $\alpha=5\%$ when increasing $k$, if $\mathscr R=\mathscr C^c$ or $\hat{\mathscr E}^c$. Also, $f_{\mathscr R}$ is reduced for $\mathscr R=\mathscr S^c$. Overall, the approximation is adequate for $h_1$ of about $30$ to $50$.
Note that $f_{\mathscr R}$ is overestimated for $h_1$ small ($k\in\{0,1\}$):
intuitively, consider $N(\mu,\sigma^2)$ RVs, such that $E_{h,t}^{(n)}\sim t(v)$ and $t(v)$ 
is heavy-tailed (especially for $n$ small), while its limit is $N(0,1)$. Larger $h_1$ may also help decrease susceptibility to outliers.

\begin{figure}[h!]
	\centering  
	\includegraphics[width=0.38\textwidth,angle=0]{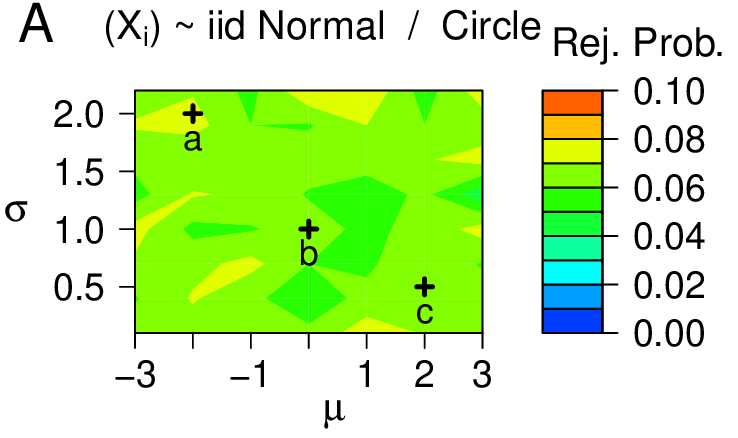}
	\includegraphics[width=0.28\textwidth,angle=0]{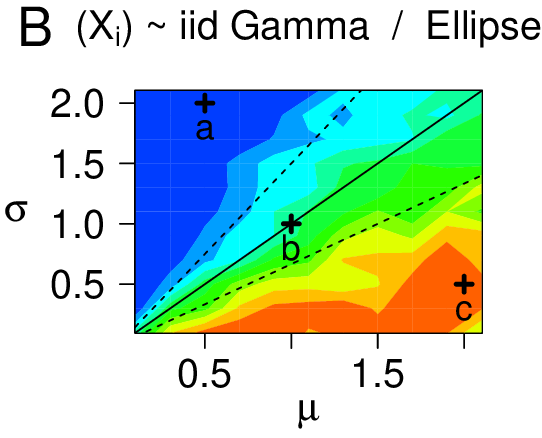}
	\includegraphics[width=0.28\textwidth,angle=0]{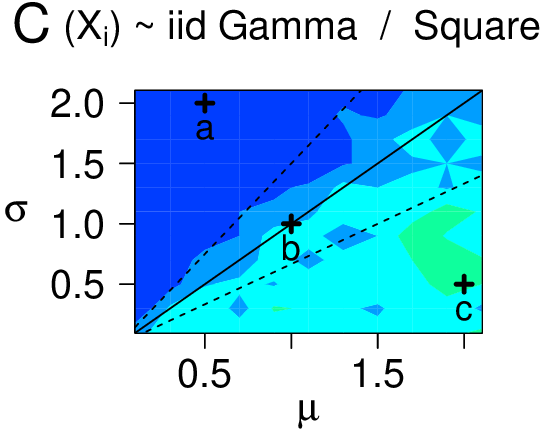}\\
	\hspace{2em}
	\includegraphics[width=0.22\textwidth,angle=0]{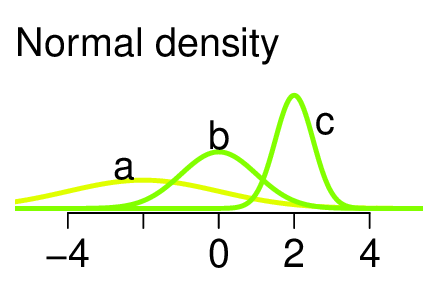}
	\hspace{7em}
	\includegraphics[width=0.22\textwidth,angle=0]{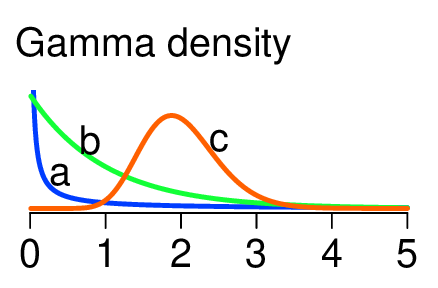}
	\hspace{2em}
	\includegraphics[width=0.22\textwidth,angle=0]{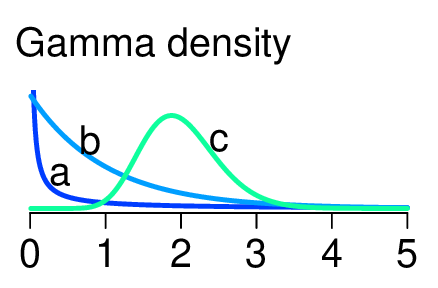}
	\caption{Top: $f_\mathscr R$ ($1000$ simulations), for $C=\emptyset$, $\alpha =5\%$, depending on $\mu$, $\sigma$, and $\mathscr R$. It is $T=1000$ and $H=\{50,75,100,125,150\}$. A: $N(\mu,\sigma^2)$ with $\mathscr R=\mathscr C^c$. B,C: $gamma(s,\lambda)$, with $\mathscr R=\mathscr E^c$ (B) and $\mathscr R=\mathscr S^c$ (C). Lines $\mu=s\cdot\sigma$ for fixed $s=1$ (solid), $s=3/2$ (lower dotted) and $s=2/3$ (upper dotted). Color-code for $f_\mathscr R$ in legend (A).
		Bottom: densities for $(\mu,\sigma)$ at '+' in top panel (B and C coincide).}
	\label{figrej2}
\end{figure}

Figure \ref{figrej2} shows the dependence of $f_\mathscr{R}$ on $\mu$ and $\sigma$. Again we see that $\alpha$ is kept for $h_1\approx 50$. We fix $H=\{50,75,100,125,150\}$ and consider $N(\mu,\sigma^2)$ with $\mathscr R=\mathscr C^c$ (A), and $gamma(s,\lambda)$ with $\mathscr R=\hat{\mathscr E}^c$ (B) or $\mathscr R=\mathscr S^c$ (C), mentioning  $s=\mu^2/\sigma^2$ and  $\lambda = \mu/\sigma^2$. The legend in A shows the color-coding of $f_{\mathscr R}$: green indicates $5\%$, red an increase and blue a decrease.


For $N(\mu,\sigma^2)$ we choose $\mu\in\{-3,-2,\ldots,3\}$ and $\sigma\in\{0.1,0.4,0.7,\ldots,2.2\}$ ($7\cdot 8 = 56$ combinations) and find $f_{\mathscr R}\approx 5\%$ for most combinations (entirely green).
 For $gamma(s,\lambda)$ we use $\mu,\sigma$ $\in\{0.1,0.3,0.5,\ldots,2.1\}$ ($11^2=121$ combinations). In B, $f_{\mathscr R}$ varies with $\mu$ and $\sigma$. For $\mu=\sigma$ (diagonal line, $s=1$, $exp(\lambda)$), it is $f_\mathscr R\approx 5\%$, which is decreased for $\mu<\sigma$, and increased for $\mu>\sigma$, but with $f_{\mathscr R}<10\%$ throughout.
 The figure suggests $f_{\mathscr R}$ to be constant for fixed $s$ and varying $\lambda$, i.e., along the lines $\mu=s\cdot\sigma$ (dotted lines, upper $s=3/4$, lower $s=3/2$), for which we mention rescaling in $J_{h,t}$ as well as the scaling property $r\cdot gamma(s,\lambda) = gamma(s,\lambda/r)$.
 Note that in B due to $\mathscr R=\hat{\mathscr E}^c$ the estimator $\hat\Gamma$ increases variability. The pattern in C is similar to B as the same distributions are considered, but overall it is more blue with $f_\mathscr{R}< 3.7\%$ throughout, aligning with the conservative nature of  $\mathscr R=\mathscr S^c$. 

\begin{figure}[h!]
	\centering  
	\includegraphics[width=0.22\textwidth,angle=0]{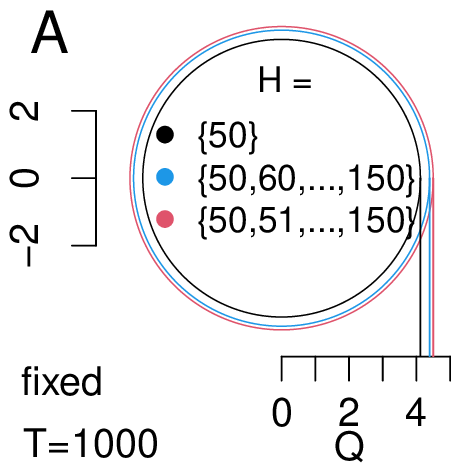}
	\hspace{1em}
	\includegraphics[width=0.22\textwidth,angle=0]{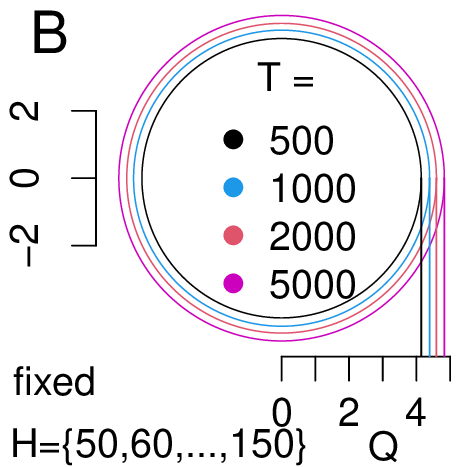}
	\caption{Dependence of $Q$ on $H$ for $T=1000$ fix (A), and on $T$ for $H=\{50,60,\ldots,150\}$ fix (B).
		It is $\alpha =5\%$, $10^6$ simulations for $Q$. Both blue circles coincide.}
	\label{fig29}
\end{figure}

\paragraph{Window choice}
We comment on the choice of $H$. For that we evaluate the dependence of $Q$ on $H$, recall (\ref{M}).
The radii in Figure \ref{fig29} show simulations of $Q$ (also denoted $Q$, $10^6$ simulations).


Figure \ref{fig29}A shows three window sets for $T=1000$ fix: first $H_1= \{50\}$ yields $Q_1\approx4.12$ (black), second $H_2=\{50,60,70,\ldots,150\}$ ($|H_2|=11$) gives $Q_2\approx 4.39$ (blue) and third $H_3=\{50,51,52,\ldots,150\}$ ($|H_3|=101$) yields $Q_3\approx 4.5$ (red). We mention monotonicity: $H_A\subset H_B$ implies $Q_A < Q_B$ which follows from global maximization over all $h\in H$, see (\ref{M}). But we also observe a flattening in the increase, i.e., $Q_2-Q_1$ (black to blue) is larger than $Q_3-Q_2$ (blue to red) although there are much more $h$ added when going from blue to red. The reason is that in $(\ref{M})$, over all $h\in H$ the $(\mathcal L_{h,t})_t$ rely on a single Brownian motion such that $\mathcal L_{h_a,t}\approx\mathcal L_{h_b,t}$ for $h_a\approx h_b$ with equality if $h_a=h_b$. Thus, if $H$ is a fine grid then additional $h$ have minor impact on the global maximum. Figure \ref{fig29}B shows four choices of $T$ for fixed $H=\{50,60,\ldots,150\}$. $T_1=500$ yields $Q_1\approx 4.14$ (black), $T_2=1000$ results in $Q_2\approx 4.39$ (blue - same setup as in A), $T_3=2000$ yields $Q_3\approx 4.6$ (red) and $T_4=5000$ results in $Q_4\approx 4.83$ (magenta). Again, we mention monotonicity: $T_A < T_B$ implies $Q_A < Q_B$. This is because in (\ref{M}) the $(\mathcal L_{h,t})_t$ are evaluated for $t\in[h,T-h]$ i.e., for larger $T$ the maxima are taken over longer intervals.

 For the choice of $H$ we now argue that first, the richer $H$ the better a scenario of unknown change points and effects is exploited: for $c\in C$ we find an $h$ preferably large but with $(c-h,c+h]$ free from other change points, resulting in an unbiased excursion of $(J_{h,t})_t$, see Propositions \ref{prop:jhc} and \ref{prop_main2}. The order of the excursion is $\sqrt{h}$, while the competitor $Q$ is bounded (for fixed $T$ and $\alpha$) by the quantile associated with all possible windows. Second, the smallest window should be chosen large enough, e.g., $50$, for the $\alpha$-level of the test to be approximately kept.

\subsection{Performance evaluation}\label{subsect:performance}
\begin{figure}[h!]
	\centering  
	\includegraphics[width=0.5\textwidth,angle=0]{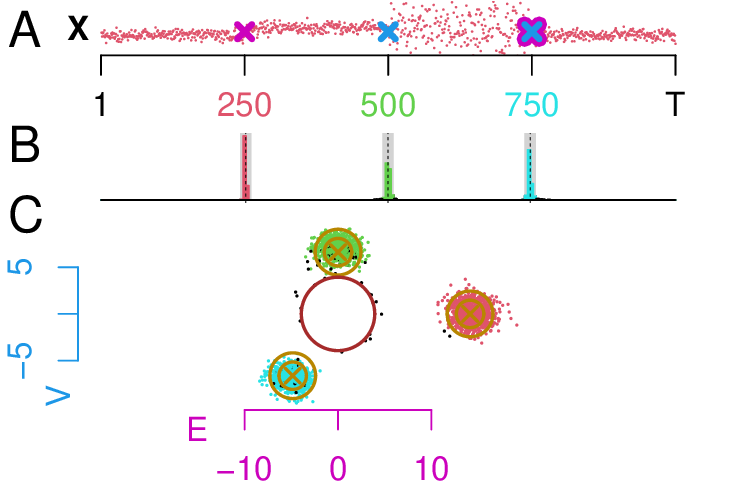}
	\includegraphics[width=0.5\textwidth,angle=0]{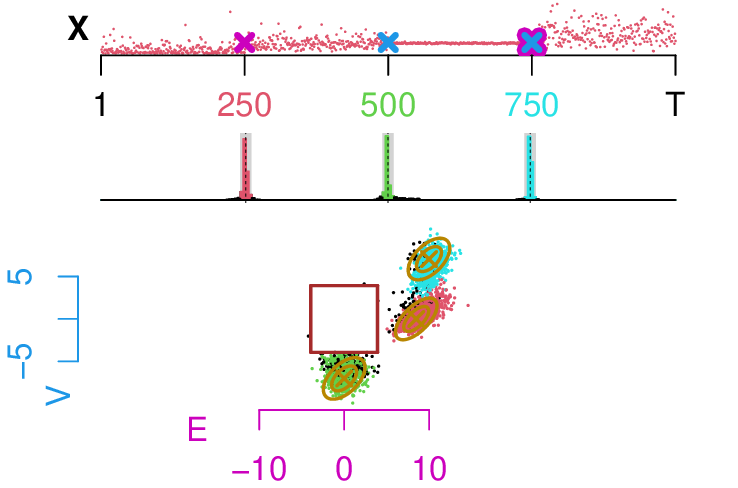}
	\begin{footnotesize}
		\begin{tabular}{ | c | c | c | c | c |}
			\hline
			& $|\hat C|$ &  \cellcolor{OrangeRed}{$|\hat \kappa_1^{(d)}|$} & \cellcolor{SeaGreen}{$|\hat \kappa_2^{(d)}|$} & \cellcolor{SkyBlue}{$|\hat \kappa_3^{(d)}|$} \\ 
			\hline
			\hline
			left 	& 3019   & \cellcolor{lightgray}{998 (1000, 988)} & \cellcolor{lightgray}{948 (998, 864)}  & \cellcolor{lightgray}{946 (989, 852)} \\ \hline
			right	& 2993   & \cellcolor{lightgray}{926 (989, 817)} & \cellcolor{lightgray}{815 (911, 720)}  & \cellcolor{lightgray}{962 (996, 913)} \\
			\hline
			\hline
		\end{tabular}
	\end{footnotesize}
	\caption{Performance evaluation. $1000$ simulations, $T=1000$, $H=\{100\}$, $C=\{250,500,750\}$ and $\alpha=5\%$. Left: $N(\mu,\sigma^2)$ with $\mu=2,10,10,2$, $\sigma=4,4,16,4$, $\mathscr R =\mathscr  C^c$. Right: $gamma(s,\lambda)$ with $\mu=0.8,2,2,4$, $\sigma=1,1,0.1,2$, $\mathscr R=\mathscr S^c$. A: $\mathbf X$, B: Distribution of all $\hat c$, color-coded: $\hat c$ $\in c_1\pm 10$ red, $\in c_2\pm 10$ green, $\in c_3\pm 10$ light blue, else black. C: $J_{h,\hat c}$ (points) colored to match $\hat c$ in B, golden: asymptotic distribution of $J_{h,c}$ for $c\in C$, brown: $\mathscr R$-boundaries. Table: $|\hat C|$ and $|\hat\kappa_u^{(10)}|,(|\hat\kappa_u^{(25)}|,|\hat\kappa_u^{(5)}|)$ for $u=1,2,3$.}
	\label{alt}
\end{figure}

\begin{figure}[h!]
	\centering  
	\includegraphics[width=0.5\textwidth,angle=0]{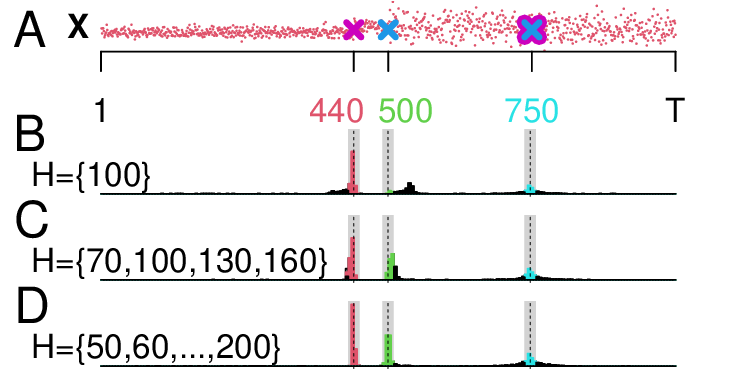}
	\includegraphics[width=0.5\textwidth,angle=0]{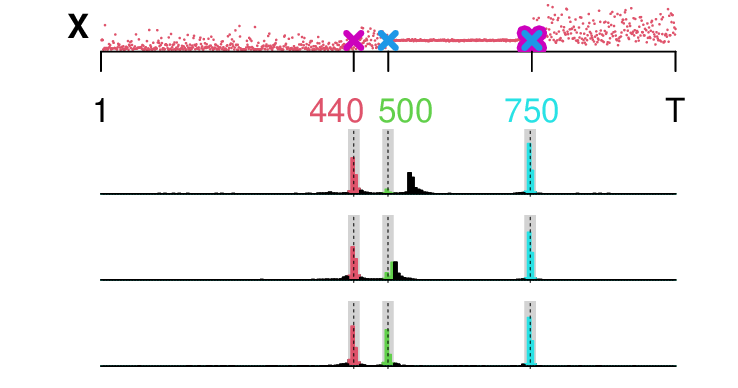}
	\begin{footnotesize}
		\begin{tabular}{ | c | c | c | c | c |}
			\hline
			& $|\hat C|$ & \cellcolor{OrangeRed}{$|\hat\kappa_1^{(d)}|$} & \cellcolor{SeaGreen}{$|\hat\kappa_2^{(d)}|$} & \cellcolor{SkyBlue}{$|\hat\kappa_3^{(d)}|$} \\ 
			\hline
			\hline
			left B	& 2040   & \cellcolor{lightgray}{778 (903, 662)} & \cellcolor{lightgray}{71 (160, 43)}  & \cellcolor{lightgray}{270 (398, 191)} \\ \hline
			left C	& 2648 & \cellcolor{lightgray}{953 (999, 631)} & \cellcolor{lightgray}{677 (894,353)} & \cellcolor{lightgray}{375 (538, 254)}   \\ \hline
			left D	& 2735 & \cellcolor{lightgray}{1000 (1000, 984)}& \cellcolor{lightgray}{872 (924, 779)} & \cellcolor{lightgray}{394 (583, 271)} \\ \hline
			\hline
			right B	& 2773 & \cellcolor{lightgray}{786 (874, 697)} & \cellcolor{lightgray}{97 (123, 77)} & \cellcolor{lightgray}{969 (998, 928)}  \\ \hline
			right C	& 2786 & \cellcolor{lightgray}{795 (927, 684)} & \cellcolor{lightgray}{367 (728, 138)} & \cellcolor{lightgray}{980 (999, 938)} \\ \hline
			right D & 2775 & \cellcolor{lightgray}{861 (959, 745)} & \cellcolor{lightgray}{676 (740, 607)} & \cellcolor{lightgray}{985 (999, 928)} \\ \hline
			\hline
		\end{tabular}
	\end{footnotesize}
	\caption{Performance evaluation. $1000$ simulations, $T=1000$, $C=\{440,500,750\}$, $\alpha=5\%$, for $H=\{100\}$ (B), $H=\{70,100,130,160\}$ (C), and $H=\{50,60,\ldots,200\}$ (D). 
		Left: $N(\mu,\sigma^2)$ with $\mu=2,10,10,6$, $\sigma=4,4,12,10$, $\mathscr R =\mathscr  C^c$. Right: $gamma(s,\lambda)$ with $\mu=0.8,2,2,4$, $\sigma=1,1,0.1,2$, $\mathscr R=\mathscr S^c$. A: $\mathbf X$. B, C, D: Distribution of all $\hat c$, color-coded: $\hat c$ $\in250\pm 25$ red, $\in500\pm 25$ green, $\in750\pm 25$ light blue, else black. Table: $|\hat C|$ and $|\hat\kappa_u^{(10)}|,(|\hat\kappa_u^{(25)}|,|\hat\kappa_u^{(5)}|)$ for $u=1,2,3$.}
	\label{fig32}
\end{figure}

For $C\not = \emptyset $ we evaluate the detection performance in simulations. We show precise estimation of the number and location of $C$, and appropriate interpretation of effects. First, we consider well-disposed setups using a single $h$, see Figure \ref{alt}. Then we reduce effects and show improvement for multiple $H$, see Figure \ref{fig32}. Finally, we consider more complex processes, see Table \ref{tab:fivecp}. 
We set $\hat\kappa_u^{(d)}:=\{\hat c\in \hat C : |\hat c-c_u|\le d\}$, 
describing 'correct' estimates of $c_u$ at distance $\le d$, for $d=10$ default, and $25$ or $5$. We perform $1000$ simulations with $T=1000$ and $\alpha=5\%$ throughout.

In Figure \ref{alt} and \ref{fig32} we differentiate $N(\mu,\sigma^2)$ with $\mathscr R=\mathscr C^c$ (left) and $gamma(s,\lambda)$ with $\mathscr R=\mathscr S^c$ (right). We consider piecewise i.i.d.~RVs. It is $C=\{c_1,c_2,c_3\}$ with a change only in $\mu$ at $c_1$ (magenta cross), a change only in $\sigma^2$ at $c_2$ (blue cross), and change in both at $c_3$ (magenta and blue cross), see $\mathbf X$ in A. $1000$ simulations yield $3000$ change points. The $10$-neighborhood of any $c_u$ is accentuated by a gray box. $\hat\kappa_u^{(10)}$ are
 color-coded according to $c_u$ ($c_1=$ red, $c_2=$ green, $c_3=$ light blue, incorrect $\hat C \backslash \cup_{u=1}^3 \hat \kappa_u^{(10)} =$ black).


In Figure \ref{alt} we set $H=\{100\}$, $C=\{250,500,750\}$, for $N(\mu,\sigma^2)$ it is $\mu=2,10,10,2$ and $\sigma=4,4,16,4$, and for $gamma(s,\lambda)$ it is $\mu=0.8,2,2,4$ and $\sigma=1,1,0.1,2$. B shows $\hat C$: three narrow histograms around the true $c_u$ support precise estimation (frequencies reported in table), e.g., left $|\hat C | =3019$. Of thousand $c_1$ it is $|\hat\kappa_1^{(d)}| =998, (1000,988)$ for $d= 10,(25,5)$, i.e., $|\hat\kappa_1^{(10)}|=998$ estimates yield the red, $|\hat\kappa_2^{(10)}|=948$ the green, $|\hat\kappa_3^{(10)}|=946$ the light blue histogram. 
It is $|\hat C \backslash \cup_{u=1}^3 \hat \kappa_u^{(10)}|=127$  (black, low number hardly visible). Panel C shows $J_{h,\hat c}$ for all $\hat c\in \hat C$, 
e.g., left $998$ red points. The $J_{h,\hat c}$ distribute closely to the asymptotic distribution of $J_{h,c}$ for $c\in C$ (golden dartboards). Thus, interpretation of effects based on $J_{h,\hat c}$ is plausible: e.g., a typical red point indicates a change only in $\mu$. We see $127$ black points: those close to the brown rejection boundary represent real false positives resulting from chance. In contrast, those in the area of the dartboards refer to true $c_u$ but are classified incorrect due to $d=10$ (tails of the histograms in B).


Before, we showed strong performance under a well-disposed setup for a single $H=\{100\}$. We now show improvement for richer $H$. We choose closer distances of the first two change points $C=\{440,500,750\}$, see Figure \ref{fig32}, and also reduce effects for $N(\mu,\sigma^2)$ to $\mu=2,10,10,6$ and $\sigma=4,4,12,10$. For $gamma(s,\lambda)$ parameters are as before. We consider three sets $H_1=\{100\}$ (B), $H_2=\{70,100,130,160\}$ (C) and $H_3=\{50,60,\ldots,200\}$ (D). Richer $H$ almost always improve the performance: first, $|\hat C|$ increases. Second, the $|\hat \kappa_u^{(d)}|$ increase. Third, B,C and D show improvement in location precision as the $\hat c$ lie closer to the true $c_u$. The small $h$ capture the rapid $c_1$ and $c_2$, and the larger $h$ improve detection of the small effect at $c_3$.
 
We further support the quality of the performance now considering five change points and different distributional assumptions, including processes with higher varying moments, see Table \ref{tab:fivecp}. We use $H=\{50,60,\ldots,200\}$ and set $C=\{200,260,500,720,810\}$ with $\mu=11,13,10,8,5,5$, $\sigma=1,3,3,3,4,1.3$, i.e., changes in $(\mu\;\&\; \sigma^2)$, $\mu$, $\mu$, $(\mu\;\&\; \sigma^2)$ and $\sigma^2$. We consider $N(\mu,\sigma^2)$ with $\mathscr R =\mathscr  C^c$, $gamma(s,\lambda)$ with $\mathscr R =\mathscr  E^c$, $unif(a,b)$ as well as for $p=2,6,10$ the periodic processes as in Section~\ref{sect:model} (mix of uniforms) with $\mathscr R =\mathscr  C^c$. Indeed, the estimates $|\hat C|$ lie close to the true $5000$, and the $|\hat\kappa_u^{(d)}|$ often approach $1000$. 


\begin{table}[h!]
	\begin{footnotesize}	
		\begin{tabular}{ | c | c | c | c | c | c | c |}
			\hline
			& $|\hat C|$ & $|\hat\kappa_1^{(d)}|$ & $|\hat\kappa_2^{(d)}|$ & $|\hat\kappa_3^{(d)}|$ & $|\hat\kappa_4^{(d)}|$ & $|\hat\kappa_5^{(d)}|$ \\ 
			\hline
			\hline
			$normal$	& 4963   & 957 (996, 853) & 845 (919, 705)  & 698 (883, 531) & 854 (974, 687) & 943 (994, 841) \\ \hline
			$gamma$	& 4598 & 970 (999, 854) & 853 (917, 710)  & 662 (879, 513) & 811 (935, 687) & 476 (584, 381) \\ \hline
			$unif$ & 4908 & 989 (1000, 916) & 844 (906, 706)  & 707 (893, 536) & 869 (953, 713) & 987 (1000, 956) \\ \hline
			$p=2$	& 4953 & 979 (1000, 920) & 840 (922, 702)  & 682 (879, 509) & 867 (977, 711) & 973 (999, 899) \\ \hline
			$p=6$ & 4958 & 981 (1000, 899) & 863 (931, 731) & 701 (884, 551) & 856 (975, 691) & 973 (998, 890) \\ \hline
			$p=10$ & 4944 & 988 (1000, 904) & 856 (928, 744) & 686 (884, 561) & 872 (973, 755) & 971 (1000, 905) \\ \hline
			\hline
		\end{tabular}
	\end{footnotesize}
	\caption{Performance evaluation. $1000$ simulations, $T=1000$, $C=\{200,260,500,720,810\}$, $\alpha=5\%$, $H=\{50,60,\ldots,200\}$, 
		$\mu=11,13,10,8,5,5$, $\sigma=1,3,3,3,4,1.3$. $N(\mu,\sigma^2)$ with $\mathscr R =\mathscr  C^c$, $gamma(s,\lambda)$ with $\mathscr R =\mathscr  E^c$, $unif(a,b)$ and for $p=2,6,10$ periodic processes as in Section~\ref{sect:model} with $\mathscr R =\mathscr  C^c$.
		Table: $|\hat C|$ and $|\hat\kappa_u^{(10)}|,(|\hat\kappa_u^{(25)}|,|\hat\kappa_u^{(5)}|)$ for $u=1,\ldots,5$.}
	\label{tab:fivecp}
\end{table}

There are other prominent methods available on \texttt{CRAN}, e.g., \texttt{mosum} \citep{CRANmosum},  \texttt{wbs} \citep{CRANwbs}, \texttt{changepoint} \citep{CRANchangepoint}, \texttt{stepR} \citep{CRANstepR}, \texttt{cumSeg} \citep{CRANcumSeg} or \texttt{FDRSeg} \citep{CRANFDRSeg}. But none of them captures all aspects covered by $\texttt{jcp}$, i.e., changes in both $\mu$ and $\sigma^2$, multiple time scales, and nonparametric methodology allowing piecewise different distributions, e.g., \texttt{mosum} essentially uses $E_{h,t}$ to address $\mu$ with a single $h$.

\subsection{Data example}\label{subsect:data}
\begin{figure}[h!]
	\centering  
	\includegraphics[width=0.6\textwidth,angle=0]{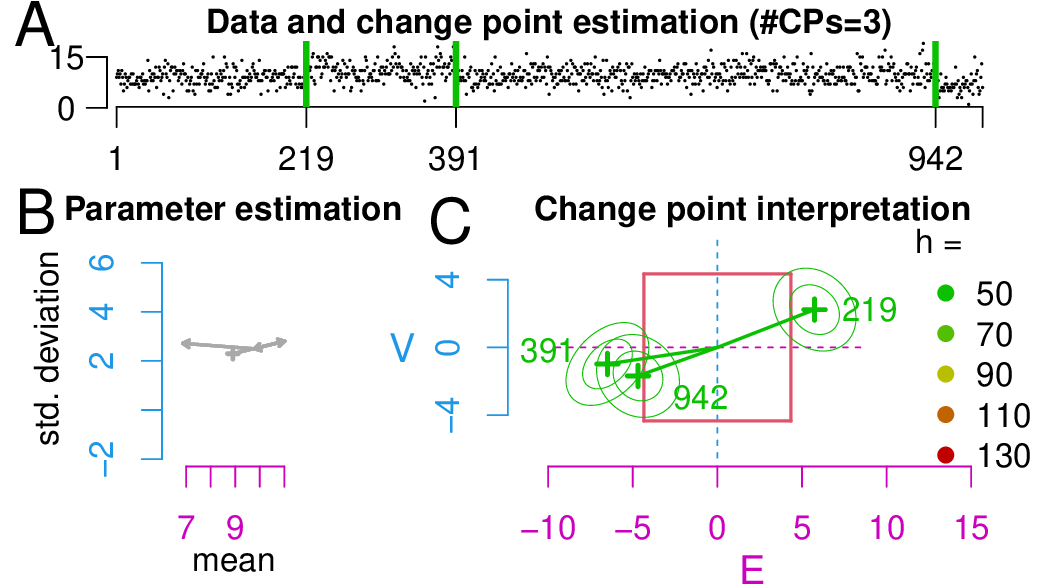}
	\caption{Frequency (in sections of length $30$) of the base uracil in the SARS-CoV-2 genome. It is $\alpha=0.05$, $\mathscr R=\mathscr S^c$ and $H=\{50,70,90,110,130\}$, yielding $\hat C=\{219,391,942\}$.}
	\label{fig35}
\end{figure}

We analyze the frequency of the nucleobase \textit{uracil} in the genome of the \textit{severe acute respiratory syndrome coronavirus 2 isolate Wuhan-Hu-1} (SARS-CoV-2), \citet[GenBank: MN908947.3]{Wu2020}. The genome contains a sequence of $29903$ bases, which we decompose into $T=996$ subsequent sections of length $30$, in each of which we compute the frequency of \textit{uracil}, see Figure \ref{fig35}A.  


We set $\alpha=0.05$, $\mathscr R=\mathscr S^c$ and $H=\{50,70,90,110,130\}$. $\mathcal H_0$ was rejected. Three change points were detected $\hat C=\{219,391,942\}$.  Figure \ref{fig35}C indicates a decrease in $\mu$ at $391$ and $942$, and an increase in $\mu$ and also slightly in $\sigma^2$ at $219$. The periodicity captured in the model (see Figure \ref{fig7}) helps mimic the data as compared to i.i.d.~sequences (see e.g.~Figure \ref{fig1}). None of the four segments shows serial correlation. Other decomposition lengths (e.g.,~$20$ or $40$) yield analogous results.  

\section{Discussion}
We proposed a method for change point detection in univariate sequences. Data are modeled as independent RVs with piecewise constant $\mu$ and $\sigma^2$, apart from which the distribution is allowed to vary periodically. In order to jointly detect changes in $\mu$ and $\sigma^2$, we developed a bivariate MOSUM approach: a comparison of means in adjacent windows is sensitive to changes in $\mu$, and a comparison of empirical variances addresses $\sigma^2$. Methodologically, first an asymptotic test was constructed to test the null hypothesis that no change occurred, $\mathcal H_0: C=\emptyset$. Second, an algorithm for change point detection was presented that can be run if $\mathcal H_0$ is rejected. 

The method is grounded on the asymptotic behavior of the MOSUM. Under $\mathcal H_0$ it was shown to approach a zero mean Gaussian process from which the rejection boundary of the test was derived in simulations. Under the alternative it asymptotically describes a Gaussian process that systematically deviates from zero locally around a change point. The quantification of this deviation supported the sensitivity of the MOSUM process to change points. The first component was shown to be sensitive to changes in $\mu$ and robust against higher order changes. The second component was shown to be sensitive to changes in $\sigma^2$, but unfortunately it also reacts to changes in $\mu$. Therefore, it is the joint consideration of both components that supports adequate estimation. Inference on the strength of a change (effect size) as well as its type ($\mu$, $\sigma^2$, or both) is enabled with confidence. 

The bivariate MOSUM was classified according to the presence or absence of symmetry (more precisely the averaged third centered moment) of the underlying distributions. Symmetry, on the one hand, was shown to result in independent asymptotic components and is thus related to Euklid's notion of distance: under $\mathcal H_0$, symmetry results first in the isotropy of the MOSUM process and thus in a rejection area given by a circle. Second, in case of change points, we found the marginals at the change points to be asymptotically uncorrelated, depicted by round contour lines. On the other hand, a lack of symmetry of the underlying distribution was shown to result in correlated components of the MOSUM process, and is therefore related to Mahalanobis' distance: first, under $\mathcal H_0$ the MOSUM has a preferred direction of variability along the main diagonals in $\rn^2$, which results in an elliptic rejection boundary. Second, the marginals at the change points have correlated components, resulting in elliptic contour lines. 
We presented three ways to treat the unknown correlation in applications: first, symmetry assumption results in vanishing correlation. Second, correlation can be estimated consistently. Third, a conservative approach that avoids to address correlation was proposed.

The method was further extended to improve the detection of change points on different scales. For that, multiple bivariate MOSUM processes were applied simultaneously. Indeed, various simulation studies revealed strong performance under different distributional assumptions, including piecewise i.i.d.~sequences and processes with varying higher moments. Generally, weak distributional assumptions allow for a wide range of applications.
The method is implemented in the $\texttt{R}$ package $\texttt{jcp}$, which performs the test and the algorithm, summarizes the results and provides a graphical output to facilitate interpretation. 

\section*{Acknowledgements}
The author is very grateful for valuable comments by Götz Kersting, Brooks Ferebee, Ralph Neininger and Anja Nowak, and for helpful suggestions of three anonymous referees.


\begin{small}
\bibliography{lit}
\end{small}

\section*{Contact information}
\vspace{-0.5em}
Michael Messer\\
Vienna University of Technology,\\ 
Institute of Statistics and Mathematical Methods in Economics,\\ 
Wiedner Hauptstraße 8-10/105, 1040 Vienna, Austria\\
tel.: +43 -1 -58801 -10588, email.: \href{mailto:michael.messer@tuwien.ac.at}{michael.messer@tuwien.ac.at}
\vspace{-0.5em}

\section*{Appendix}\label{appendix} 
We give here all proofs and additional auxiliary results.
\paragraph{ad Section \ref{sect:evj}} First we state a functional strong law of large numbers (SLLN) for the auxiliary processes, see Definition \ref{defi:aux}.
\begin{lemm}\label{lemm_aux}
	For an auxiliary process $\textbf{X}=(X_i)_{i=1,2,\ldots}$
	of period $p$ and moments 
	$\mu^{\langle k \rangle} = (1/p)\sum_{m=1}^p  \mu(m)^{\langle k \rangle}$
	it is for $k\in \{1,\ldots,4\}$ in $(\mathcal D_\rn[0,T],d_{\|\cdot\|})$ a.s.~as $n\to\infty$ 
	\begin{align}\label{fslln}
		\left( \frac{1}{n}\sum\nolimits_{i=1}^{\lfloor nt\rfloor} X_i^k\right)_t \longrightarrow (t\mu^{\langle k \rangle})_t.
	\end{align}
\end{lemm}	
\noindent
\textbf{Proof}: 
The consecutive blocks $(X_{lp+1},\ldots,X_{(l+1)p})_{l=0,1,\ldots}$ of length $p$ constitute an i.i.d.~sequence. Define the mean $Z_{l}^{\langle k\rangle}:= (1/p)\sum_{m=1}^p X_{lp + m}^k$. Then, for $n\to\infty$ a.s.
\begin{align}\label{slln}
	\frac{1}{n}\sum\nolimits_{i=1}^n X_i^k 
	= 
	\Big(
	\frac{p}{n} \sum\nolimits_{l=0}^{\lfloor n/p \rfloor -1}
	Z_{l}^{\langle k\rangle}
	\Big)
	+ 
	\Big(
	\frac{1}{n} \sum\nolimits_{i=\lfloor n/p \rfloor p +1}^n X_i^k
	\Big)
	\longrightarrow \mu^{\langle k \rangle}.
\end{align} 
The SLLN applied to 
$(Z_{l}^{\langle k\rangle})_{l=0,1,\ldots}$
shows the first summand to tend to $\mu^{\langle k\rangle}$. The second summand is the remainder of the last block and vanishes as the moment assumptions imply $(1/n)\sum_{m=1}^p |X_{lp+m}|^k \to 0$ a.s. as $n\to\infty$. From (\ref{slln}) we obtain $\sup_{0\le t \le T}|(1/n)\sum_{i=1}^{\lfloor nt \rfloor} X_i^k- t\mu^{\langle k\rangle}|\to 0$ by discretizing time. \hfill\qed

\noindent\\
\textbf{Proof of Lemma \ref{conv:est}}: 
We aim to apply Lemma \ref{lemm_aux}, for which we express the centered population parameters from  (\ref{theo_moments}) through the non-centered moments as
\begin{align}\label{decomp}
	\mu^{\{2\}} = \frac{1}{p}\sum\nolimits_{m=1}^p \mu(m)^{\{2\}}= \frac{1}{p}\sum\nolimits_{m=1}^p \mu(m)^{\langle 2 \rangle} - \frac{1}{p}\sum\nolimits_{m=1}^p \mu(m)^{2}
	= \mu^{\langle 2 \rangle} - \mu^2,
\end{align}
and analogously $\mu^{\{3\}} =\mu^{\langle 3 \rangle} - 3\mu\mu^{\langle 2 \rangle} + \mu^3$
and $
\mu^{\{4\}} =\mu^{\langle 4 \rangle} - 4\mu\mu^{\langle 3 \rangle} + 6\mu^2\mu^{\langle 2 \rangle}  -3\mu^4$.
Note that $\mu(m)=\mu$ constant implies $(1/p)\sum_{m=1}^p \mu(m)^{2}=\mu^2$. This is favorable, as Lemma \ref{lemm_aux} only states estimation of
$\mu^{\langle k \rangle}= (1/p)\sum_{m=1}^p \mu(m)^{\langle k \rangle}=(1/p)\sum_{m=1}^p \ew[X_m^k]$, but not of
$(1/p)\sum_{m=1}^p \mu(m)^{k}=(1/p)\sum_{m=1}^p \ew[X_m]^{k}$ if $\mu(m)$ is not constant. 
Analogously, the average 
\[
\nu^{2} = \frac{1}{p}\sum\nolimits_{m=1}^p \nu(m)^{\{4\}}= \frac{1}{p}\sum\nolimits_{m=1}^p \mu(m)^{\{4\}} - \frac{1}{p}\sum\nolimits_{m=1}^p (\sigma(m)^{2})^2
= \mu^{\{4\}} - \sigma^4
\]
from (\ref{nu}) behaves well as $\sigma(m)^2=\sigma^2$ constant yields $(1/p)\sum_{m=1}^p (\sigma(m)^{2})^2 = \sigma^4$. If $\sigma(m)^2$ was not constant, 
$\hat\sigma_j^4$ would fail in estimating $(1/p)\sum_{m=1}^p (\sigma(m)^{2})^2$.\\
We consider $j=\ell$. Note that $\hat\mu_\ell^{\{k\}}$ from (\ref{moments}) can be expressed through $\hat\mu_\ell^{\langle k\rangle}$ as
$
\hat \mu_\ell^{\{2\}} = \hat\mu_\ell^{\langle 2 \rangle} - \hat\mu_\ell^2$,
$\hat \mu_\ell^{\{3\}}  =\hat\mu_\ell^{\langle 3 \rangle} - 3\hat\mu_\ell\hat\mu_\ell^{\langle 2 \rangle} + \hat\mu_\ell^3$ and $
\hat\mu_\ell^{\{4\}} =\hat\mu_\ell^{\langle 4 \rangle} - 4\hat\mu_\ell\hat\mu_\ell^{\langle 3 \rangle} + 6\hat\mu_\ell^2\hat\mu_\ell^{\langle 2 \rangle}  -3\hat \mu_\ell^4$.
A comparison with (\ref{decomp}) reveals that it suffices
to have
$(\hat{\mu}_\ell^{\langle k \rangle})_t\to(\mu^{\langle k\rangle})_t$  a.s.~in $(\mathcal D_\rn [h,T-h],d_{\|\cdot\|})$, which we obtain by Lemma \ref{lemm_aux} as for $n\to\infty$
\begin{small}
	\begin{align*}
		(\hat{\mu}_\ell^{\langle k \rangle})_t
		= 
		\Big( \frac{1}{nh}
		\sum\limits_{i=1}^{\lfloor nt\rfloor} X_i^k\Big)_t
		- \Big( \frac{1}{nh}
		\sum\limits_{i=1}^{\lfloor n(t-h)\rfloor} X_i^k\Big)_t
		\longrightarrow
		\Big(\frac{t }{h}\mu^{\langle k\rangle}\Big)_t - \Big(\frac{t - h }{h}\mu^{\langle k\rangle} \Big)_t = (\mu^{\langle k\rangle})_t.\qed
	\end{align*}
\end{small}
\noindent\\
In order to prepare the proof of Proposition \ref{prop_main}, we state a bivariate functional central limit theorem (CLT) for the first and second moment of an auxiliary process.

\begin{lemm}\label{lemm:fclt}
	Let $\textbf{X}=(X_i)_{i=1,2,\ldots}$ be an auxiliary process with population parameters $\mu$, $\sigma^2$, $\nu^2$ and $\rho$.  Rescale variables as $Y_i := (X_i -\mu)/\sigma$ and $\mathcal Y_i := [(X_i-\mu)^2-\sigma^2]/\nu$ and let $\Gamma$ be as in (\ref{gamma}). 
	Then it holds in $(\mathcal D_{\rn^2}[0,\infty),d_{SK})$ as $n\to\infty$ 
	\begin{align}\label{donsker}
		\left[
		\frac{1}{\sqrt{n}}\;\Gamma^{-1/2}\cdot
		\begin{pmatrix}
			\sum_{i=1}^{\lfloor nt \rfloor} Y_i\\
			\sum_{i=1}^{\lfloor nt \rfloor} \mathcal Y_i
		\end{pmatrix}
		\right]_{t}
		\stackrel{d}{\longrightarrow}
		\begin{pmatrix}
			W_t\\
			\mathcal W_t
		\end{pmatrix}_{t},
	\end{align}
	while $(W_t,\mathcal W_t)^\mathsf{T}$ constitutes a planar Brownian motion.
\end{lemm}	

\noindent
\textbf{Proof}:
The blocks $(X_{lp+1},\ldots,X_{(l+1)p})_{l=0,1,\ldots}$ are i.i.d.
Set $Z_{l}:=(1/\sqrt{p})\sum_{m=1}^{p} Y_{lp+m}$ and $\mathcal Z_{l}:=(1/\sqrt{p})\sum_{m=1}^{p} \mathcal Y_{lp+m}$. Then 
$(Z_{l},\mathcal Z_l)_{l=0,1,\ldots}$ is an i.i.d.~sequence with components having zero mean and unit variance. Regarding (\ref{donsker}) we rewrite
\begin{align}\label{decomp_donsker}
	\frac{1}{\sqrt{n}}\sum\nolimits_{i=1}^{\lfloor nt \rfloor} Y_i
	=
	\Big(
	\sqrt{\frac{p}{n}}
	\sum\nolimits_{l=0}^{\lfloor (n/p)t\rfloor-1} 
	Z_l 
	\Big)
	+ 
	\Big(
	\frac{1}{\sqrt{n}}
	\sum\nolimits_{i=\lfloor (n/p)t \rfloor p +1}^{\lfloor nt \rfloor} Y_{i}
	\Big),
\end{align}
using $\lfloor nt/p\rfloor = \lfloor \lfloor nt \rfloor /p\rfloor$.

The second summand vanishes as $(1/\sqrt{n})\sum_{m=1}^p|Y_{lp+m}|\to0$ in probability as $n\to\infty$. We treat the second component of (\ref{donsker}) analogously. As $\mathbb{C}ov((Z_1,\mathcal Z_1)^{\mathsf T})=\Gamma$, the result follows from a functional CLT w.r.t.~$(Z_{l},\mathcal Z_l)_{l=0,1,\ldots}$, see \citet{Kuelbs1973} working on general Banach space valued RVs.
\hfill\qed\\

\noindent
\textbf{Proof of Proposition \ref{prop_main}}:
We define a continuous map $\varphi=(\varphi_1,\varphi_2)^\mathsf{T}$ from $(\mathcal D_{\rn^2}[0,T],d_{SK})$ to $(\mathcal D_{\rn^2}[h,T-h],d_{SK})$, for both components  $\kappa=1,2$ via
\begin{align}\label{phi}
	\varphi_\kappa : (f_\kappa(t))_{t} \to \left(\frac{[f_\kappa(t+h)-f_\kappa(t)]-[f_\kappa(t)-f_\kappa(t-h)]}{\sqrt{2h}}\right)_t.
\end{align}
Applying $\varphi$ on (\ref{donsker}) yields in $(\mathcal D_{\rn^2}[h,T-h],d_{SK})$ as $n\to\infty$ 
\begin{align*}
	\left[
	\Gamma^{-1/2}\cdot
	\begin{pmatrix}
		(2\sigma^2nh)^{-1/2}
		\left[\sum_{i\in I_r} X_i - \sum_{i\in I_\ell} X_i\right]\\ 
		(2\nu^2nh)^{-1/2}
		\left[\sum_{i\in I_r} (X_i-\mu)^2 - \sum_{i\in I_\ell} (X_i-\mu)^2\right]
	\end{pmatrix}
	\right]_{t}
	\stackrel{d}{\longrightarrow}
	\mathcal (\mathcal L_{h,t})_t.
\end{align*}
The centering $\mu$ and $\sigma^2$ canceled. In the second component replace $\mu$ with $\hat\mu_j$ yielding
\begin{align}\label{conv:trueorder}
	\left[
	\Gamma^{-1/2}\cdot
	\begin{pmatrix}
		(\hat{\mu}_r -\hat{\mu}_\ell)/\sqrt{2\sigma^2/(nh)}\\
		(\hat\sigma_r^2 -\hat\sigma_\ell^2)/\sqrt{2\nu^2/(nh)}
	\end{pmatrix}
	\right]_{t}
	\stackrel{d}{\longrightarrow}
	\mathcal (\mathcal L_{h,t})_t.
\end{align}
Regarding the replacement we note that for $j\in\{\ell,r\}$
\begin{align*}
	\frac{1}{\sqrt{nh}}\sum\nolimits_{i\in I_j} (X_i - \mu)^2
	= 
	\frac{1}{\sqrt{nh}}\sum\nolimits_{i\in I_j} (X_i - \hat\mu_j)^2
	-
	\Big(
	(\hat\mu_j-\mu)\frac{1}{\sqrt{nh}}\sum\nolimits_{i\in I_j} (X_i - \mu)
	\Big),
\end{align*}
while the second summand vanishes uniformly over $t$ a.s.~as $n\to\infty$, as first Lemma \ref{conv:est} states $(\hat\mu_j)_t \to (\mu)_t$ a.s., and second square-root scaling in $[(nh)^{-1/2}\sum_{i\in I_j} (X_i - \mu)]_t$ yields weak convergence to a Gaussian process limit.
Finally, Lemma \ref{conv:est} also implies that $2\sigma^2$ and $2\nu^2$ in (\ref{conv:trueorder}) can be replaced by $\hat\sigma_r^2 + \hat\sigma_\ell^2$ and $\hat\nu_r^2 + \hat\nu_\ell^2$. \hfill\qed

\paragraph{ad Section \ref{sect:test}}$\,$\\
\noindent
\textbf{Proof of Theorem \ref{theo_main}}: The result for $M_h^{(n)}(\Gamma)$ follows from Proposition \ref{prop_main} applying continuous mapping. From this, the statements for $M_h^{(n)}(I)$ and $M_h^{(n)}(\infty)$ follow by construction. For $M_h^{(n)}(\hat \Gamma)$ we use consistency of $\hat \rho$, see Lemma \ref{conv:est}. \hfill\qed

\paragraph{ad Section \ref{sect:cpd}}$\,$\\
\noindent
\textbf{Proof of Proposition \ref{prop:jhc}}: 
We write 
$
J_{h,c}^{(n)} - \Delta_{h,c}^{(n)} \cdot  j_{h,c}^{(n)} = \Delta_{h,c}^{(n)} \cdot D_{h,c}^{(n)}
$
with 
\begin{align}\label{dhc}
	D_{h,c}^{(n)}:=
	\sqrt{nh}\cdot
	\begin{pmatrix}
		\dfrac{(\hat{\mu}_r-\hat{\mu}_\ell) - (\mu_2-\mu_1)}{(\sigma_2^2 + \sigma_1^2)^{1/2}},
		\dfrac{(\hat\sigma_2^2-\hat\sigma_1^2) - (\sigma_2^2-\sigma_1^2)}{(\nu_2^2 + \nu_1^2)^{1/2}}\\
	\end{pmatrix}^\mathsf{T}.
\end{align}
We need to show $D_{h,c}^{(n)}\stackrel{d}{\longrightarrow} N_2(0,\Gamma_c)$ as $n\to\infty$, as $\Delta_{h,c}^{(n)}\to I$ a.s.~by components, see Lemma \ref{conv:est}. 
Set $\hat\theta_{h,c}^{(n)}:= (\hat{\mu}_r,\hat\sigma_r^2,\hat{\mu}_\ell,\hat\sigma_\ell^2)^\mathsf{T}$ and 
$\theta:= (\mu_2,\sigma_2^2,\mu_1,\sigma_1^2)^\mathsf{T}$. As $n\to\infty$
\begin{align*}
	\sqrt{nh}\cdot
	(\hat\theta_{h,c}^{(n)}-\theta)
	\stackrel{d}{\longrightarrow} 
	N_4(0,\Sigma) 
	\quad\textrm{with}\quad
	\Sigma = 
	\begin{pmatrix}
		\sigma_2^2 & \mu_2^{\{3\}} & 0 & 0\\
		\mu_2^{\{3\}}  & \nu_2^2 & 0 & 0\\
		0& 0 & \sigma_1^2 &  \mu_1^{\{3\}} \\
		0 &0 &  \mu_1^{\{3\}}  & \nu_1^2
	\end{pmatrix},
\end{align*}
using the Lindeberg-Feller CLT in each component while periodicity of the RVs implies the Lindeberg condition.
Cram\'er-Wold device yields joint convergence. 
Set
\[
A := 
\begin{pmatrix}
	1 & 0 & -1 & 0\\
	0 & 1 & 0 & -1
\end{pmatrix}
\qquad \textrm{and}\qquad
\Sigma^* := A\Sigma A^{\mathsf{T}} = 
\begin{pmatrix}
	\sigma_2^2 + \sigma_1^2 & \mu_2^{\{3\}} + \mu_1^{\{3\}}\\
	\mu_2^{\{3\}} + \mu_1^{\{3\}} & \nu_2^2 + \nu_1^2
\end{pmatrix}.
\]
Then, the continuous mapping theorem yields as $n\to\infty$
\[ 
\sqrt{nh}
\begin{pmatrix}
	(\hat{\mu}_r - \hat{\mu}_\ell) - (\mu_2-\mu_1)\\
	(\hat\sigma_r^2 - \hat\sigma_\ell^2) - (\sigma_2^2 -\sigma_1^2)
\end{pmatrix}
= A\cdot [\sqrt{nh}\cdot (\hat\theta_{h,c}^{(n)}-\theta)] 
\stackrel{d}{\longrightarrow} 
N_2(0,\Sigma^*),
\]
and further
\begin{align*}
	D_{h,c}^{(n)} & =
	\begin{pmatrix}
		(\sigma_2^2+\sigma_1^2)^{-1/2} & 0 \\
		0 & (\nu_2^2+\nu_1^2)^{-1/2}
	\end{pmatrix}
	\cdot A\cdot [\sqrt{nh}\cdot (\hat\theta_{h,c}^{(n)}-\theta)] 
	\stackrel{d}{\longrightarrow} N_2(0,\Gamma_c),
\end{align*}
since $\Gamma_c = D\Sigma^* D^\mathsf{T}$, while $D$ denotes the diagonal matrix of the latter display. \hfill\qed\

\paragraph{ad Section \ref{sect:jht_c}}$\,$\\
\textbf{Proof of Lemma \ref{lemm:conv_est_c}}: W.l.o.g.~let $j=\ell$. Lemma \ref{conv:est} implies pointwise a.s.~convergence to the population parameter if $(t-h,h]\not\ni c$. Else, a proportion of $[c-(t-h)]/h$ RVs belongs to $\mathbf X_1$, and $(t-c)/h$ to $\mathbf X_2$.
Regarding $\hat\mu_\ell$ we obtain a.s.~as $n\to\infty$
\begin{align}\label{tilde_mu_aux}
	\hat\mu_\ell 
	&= 
	\Big[
	\frac{nc - \lfloor  n(t-h)\rfloor}{nh}
	\frac{1}{nc - \lfloor  n(t-h)\rfloor}
	\sum\nolimits_{i=\lfloor  n(t-h)\rfloor +1}^{nc} X_{1,i}
	\Big]\\
	&\qquad \;+\; 
	\Big[
	\frac{\lfloor  nt\rfloor - nc}{nh}
	\frac{1}{\lfloor  nt\rfloor - nc}
	\sum\nolimits_{i=nc +1}^{\lfloor  nt\rfloor} X_{2,i}\Big]
	\longrightarrow 
	\frac{c-(t-h)}{h} \mu_1 + 
	\frac{t-c}{h} \mu_2.\nonumber
\end{align} 
Uniform convergence follows from Lemma \ref{lemm_aux}. Analogously, for $\hat\sigma_\ell^2$ and $\hat\nu_\ell^2$ it remains to consider $(t-h,h]\ni c$. We replace $X_{u,i}$ in (\ref{tilde_mu_aux}) and discuss the left subinterval $(t-h,c]$. For $\hat\sigma_\ell^2$ we find a.s.~as $n\to\infty$
\begin{align*}
	\frac{1}{nc - \lfloor  n(t-h)\rfloor}
	\sum\nolimits_{i=\lfloor  n(t-h)\rfloor +1}^{nc} (X_{1,i} - \hat\mu_\ell)^2 
	\longrightarrow \sigma_1^2+(\tilde\mu_\ell -\mu_1)^2.
\end{align*}
For that, decompose
$(X_{1,i}-\hat\mu_\ell)^2 = (X_{1,i}-\mu_1)^2-2(X_{1,i}-\mu_1)(\hat\mu_\ell-\mu_1) + (\hat\mu_\ell-\mu_1)^2$. The first summand yields $\sigma_1^2$, the second vanishes, and the third summand yields $(\tilde\mu_\ell -\mu_1)^2$. Use here $\hat\mu_\ell\to \tilde\mu_\ell$ a.s.
For $\hat\nu_\ell^2=\hat\mu_\ell^{\{4\}}-\hat\sigma_\ell^4$ it remains to comment on $\hat\mu_\ell^{\{4\}}$.
We find a.s.~as $n\to\infty$ 
\begin{align*}
	\frac{1}{nc - \lfloor  n(t-h)\rfloor}&
	\sum\nolimits_{i=\lfloor  n(t-h)\rfloor +1}^{nc} (X_{1,i} - \hat\mu_\ell)^4
	\\
	&\qquad\qquad
	\longrightarrow
	\mu_1^{\{4\}} - 4\mu_1^{\{3\}}(\tilde\mu_\ell-\mu_1) + 6\sigma_1^2(\tilde\mu_\ell-\mu_1)^2 + (\tilde\mu_\ell-\mu_1)^4.
\end{align*} 
For this we decompose
$(X_{1,i} - \hat\mu_\ell)^4 =
(X_{1,i} - \mu_1)^4
- 4(X_{1,i}-\mu_1)^3 (\hat\mu_\ell-\mu_1)
+6(X_{1,i}-\mu_1)^2 (\hat\mu_\ell-\mu_1)^2
-4(X_{1,i}-\mu_1) (\hat\mu_\ell-\mu_1)^3
+ (\hat\mu_\ell-\mu_1)^4,
$
using $\hat\mu_\ell\to \tilde\mu_\ell$ and $\hat\sigma_\ell^2\to \tilde\sigma_\ell^2$ a.s. The fourth summand vanishes.\hfill\qed\\
\noindent\\
\textbf{Proof of Proposition \ref{prop_main2}}:
$\mathbf{X}$ relies on  independent $\mathbf X_1$ and $\mathbf X_2$. Let $u\in\{1,2\}$. $\mathbf X_u$ has population parameters $\mu_u$, $\sigma_u^2$ and $\nu_u^2$. We standardize $Y_{u,i} := (X_{u,i} -\mu_u)/\sigma_u$ and $\mathcal Y_{u,i} := [(X_{u,i}-\mu_u)^2-\sigma_u^2]/\nu_u$. 
The functional CLT from Lemma \ref{lemm:fclt} states that for $u\in\{1,2\}$ it holds in $(\mathcal D_{\rn^2}[0,T],d_{SK})$ as $n\to\infty$ 
\begin{align}\label{donsker_u}
	\left[
	\frac{1}{\sqrt{n}}\;
	\begin{pmatrix}
		\sum_{i=1}^{\lfloor nt \rfloor} Y_{u,i}\\
		\sum_{i=1}^{\lfloor nt \rfloor} \mathcal Y_{u,i}
	\end{pmatrix}
	\right]_{t}
	\stackrel{d}{\longrightarrow}
	\begin{pmatrix}
		W_{u,t}\\
		\mathcal W_{u,t}
	\end{pmatrix}_{t},
\end{align}
while
$(W_{u,t},\mathcal W_{u,t})_t$ is a planar Brownian motion. It is $\Gamma_u=I$ as $\rho_u=0$. The independence of $\mathbf X_1$ and $\mathbf X_2$ inherits to $(W_{1,t},\mathcal W_{1,t})_t$ and $(W_{2,t},\mathcal W_{2,t})_t$. It also implies
joint convergence (\ref{donsker_u}) over $u=1$ and $2$. We replace $\sigma_u$ and $\nu_u$ with $[h(\tilde\sigma_r^2+\tilde\sigma_\ell^2)]^{1/2}$ and $[h(\tilde\nu_r^2+\tilde\nu_\ell^2)]^{1/2}$, see (\ref{tilde_vartheta}) and (\ref{tilde_left}), yielding in
$(\mathcal D_{\rn^2}[h,T-h],d_{SK})$ as $n\to\infty$ 
\begin{small}
	\begin{align}\label{donsker_u2}
		\left[
		\frac{1}{\sqrt{nh}}\; 
		\begin{pmatrix}
			[\sigma_u / (\tilde\sigma_r^2+\tilde\sigma_\ell^2)^{1/2}]
			\sum_{i=1}^{\lfloor nt \rfloor} Y_{u,i}\\
			[\nu_u / (\tilde\nu_r^2+\tilde\nu_\ell^2)^{1/2}]
			\sum_{i=1}^{\lfloor nt \rfloor} \mathcal Y_{u,i}
		\end{pmatrix}
		\right]_{t}
		\stackrel{d}{\longrightarrow}
		\left[
		\frac{1}{\sqrt{h}}
		\begin{pmatrix}
			[\sigma_u / (\tilde\sigma_r^2+\tilde\sigma_\ell^2)^{1/2}] \cdot W_{u,t}\\
			[\nu_u / (\tilde\nu_r^2+\tilde\nu_\ell^2)^{1/2}] \cdot
			\mathcal W_{u,t}
		\end{pmatrix}
		\right]_{t}.
	\end{align}
\end{small}
To switch to the moving sum perspective, we define a continuous map
$
\varphi : (\mathcal D_{\rn^2}[h,T-h]\times \mathcal D_{\rn^2}[h,T-h],d_{SK}\otimes d_{SK}) \to 
(\mathcal D_{\rn}[h,T-h]\times \mathcal D_{\rn}[h,T-h],d_{SK}\otimes d_{SK}).
$
For that we write an element from the domain of $\varphi$ as $\iota:=[((f_{1},f_{2})(t))_{t},((g_{1},g_{2})(t))_{t}]$, with $f_\kappa$ associated with the $\kappa$-th component of the first process $(u=1)$, and $g_\kappa$ with the $\kappa$-th component of the second process $(u=2)$, for $\kappa=1,2$. Define $\varphi=(\varphi_1,\varphi_2)$ componentwise for $\kappa=1,2$ identical via 
\begin{small}
	\begin{align*}
		\varphi_\kappa(\iota) :=
		&\left( \begin{array}[c]{l}
			[(f_\kappa(t+h)-f_\kappa(t)) - (f_\kappa(t)-f_\kappa(t-h))]\mathbbm{1}_{[h,c-h)}(t) 	\\
			+[(g_\kappa(t+h)-g_\kappa(c))+(f_\kappa(c)-f_\kappa(t)) - (f_\kappa(t)-f_\kappa(t-h))]\mathbbm{1}_{[c-h,c)}(t) \\
			+[(g_\kappa(t+h)-g_\kappa(t))-(g_\kappa(t)-g_\kappa(c)) - (f_\kappa(c)-f_\kappa(t-h))]\mathbbm{1}_{[c,c+h)}(t)\\
			+[(g_\kappa(t+h)-g_\kappa(t)) - (g_\kappa(t)-g_\kappa(t-h))]\mathbbm{1}_{[c+h,T-h]}(t)\\
		\end{array}\right)_{t}.	
	\end{align*}
\end{small}
We apply $\varphi_\kappa$ on (\ref{donsker_u2}). Continuous mapping preserves convergence. The first ($t< c-h$) and the fourth ($t\ge c+h$) summand in $\varphi_\kappa$ refer to a single $\mathbf X_u$ for which we obtain $(J_{h,t}^{(n)})_t\stackrel{d}{\longrightarrow} (\mathcal L_{h,t})_t$ as in Proposition \ref{prop_main}.
We need to discuss $t\in(c-h,c+h]$.
We focus on the case that $c$ lies in the right window $t\in [c-h,c)$, and consider the first component $\kappa=1$. Application of $\varphi_1$ on the left hand side of (\ref{donsker_u2}) yields
\begin{align}\label{alt_tech}
	&\sqrt{nh}
	\left[
	\frac{1}{nh}
	\left(
	\sum\nolimits_{i=\lfloor nc\rfloor +1}^{\lfloor n(t+h)\rfloor} X_{i,2}
	+
	\sum\nolimits_{i=\lfloor nt\rfloor +1}^{\lfloor nc \rfloor} X_{i,1}
	\right)
	- 
	\frac{1}{nh}
	\sum\nolimits_{i=\lfloor n(t-h)\rfloor +1}^{\lfloor nt \rfloor} X_{i,1}
	\right]\\
	&\,\,\quad
	- \sqrt{nh}
	\left[
	\frac{[((t+h)-c)\mu_2-(c-t)\mu_1]/h - \mu_1}{(\tilde\sigma_r^2+\tilde\sigma_\ell^2)^{1/2}}
	\right]
	= \sqrt{nh} 
	\Big[
	\frac{\hat\mu_r-\hat\mu_\ell}{\sqrt{\tilde\sigma_r^2+\tilde\sigma_\ell^2}} - 
	\frac{\tilde\mu_r-\tilde\mu_\ell}{\sqrt{\tilde\sigma_r^2+\tilde\sigma_\ell^2}}
	\Big].\nonumber 
\end{align}
For the right hand side of (\ref{donsker_u2}) we obtain
\begin{align}\label{limit_alt}
	\frac{\sigma_2(W_{2,t+h}-W_{2,c}) -\sigma_1[(W_{1,c}-W_{1,t})-(W_{1,t}-W_{1,t-h})]}{\sqrt{(\tilde\sigma_r^2+\tilde\sigma_\ell^2)h}},
\end{align}
which has zero expectation. Replacing $W_{1,t}$ and $W_{2,t}$ with a single Brownian motion $(W_t)_t$ yields continuity. We replace the scaling: first, multiplication of (\ref{alt_tech}) with the first entry of $\Delta_{h,t}^{(n)}$ 
preserves the limit as $(\Delta_{h,t}^{(n)})_t\to(I)_t$ a.s. Second, multiplication of both (\ref{alt_tech}) and (\ref{limit_alt}) with the first entry of $(\tilde D_{h,t})_t$ scales (\ref{limit_alt}) to unit variance. Set the latter as the first component of $(\mathcal L_{h,t}^*)_t$. In total, the first component of (\ref{conv:jht}) holds true. Analogously, derive the second component. It is $\mathcal L_{h,t}^*\sim N_2(0,I)$.
\hfill\qed

\paragraph{ad Section \ref{sect:practical}}$\,$\\
\textbf{Proof of Corollary \ref{coro_main}}: 
Set $\varphi^h:=\varphi$ as in (\ref{phi}) and proceed analogously to the proof of Proposition \ref{prop_main}
by applying the joint map $(\varphi^{h_1},\ldots,\varphi^{h_w})$ on (\ref{donsker}).\hfill\qed\\

\noindent
\textbf{Proof of Theorem \ref{theo_main2}}: 
We argue analogously to the proof of Theorem \ref{theo_main}, applying Corollary \ref{coro_main}.\hfill\qed

\end{document}